\documentclass[12pt,a4paper]{article} 
\usepackage[affil-it]{authblk}
\usepackage{amsmath}
\usepackage{bm}

\usepackage{cite}
\usepackage{pstricks}
\usepackage{graphicx}
\usepackage{subfigure}
\usepackage{caption}
\usepackage{pdflscape}
\usepackage{amssymb}
\usepackage{color}
\usepackage{url}

\begin{document}

\title{Elastic stress effects on microstructural instabilities\thanks{We dedicate this paper to the memory of John W Cahn. 
His immense contributions to the field of elastic stress induced
microstructural instabilities are seen not only in his papers, but also in a large number of  
acknowledgements that we noticed during the preparation of this manuscript -- such as the acknowledgement 
of Tien and Copley in their classic paper on rafting.}}

\author{M P Gururajan \thanks{Corresponding author: $gururajan.mp@gmail.com$,$guru.mp@iitb.ac.in$}} 

\affil{Department of Metallurgical Engineering and Materials Science, Indian Institute of Technology Bombay, 
Powai, Mumbai, 400076 INDIA}

\author{Arka Lahiri}

\affil{Department of Materials Engineering, Indian Institute of Science, Bengaluru, 560012 INDIA}

\date{}

\maketitle

Microstructure (which, for the purposes of this article, defined as the sizes, shapes and distributions of 
interfaces in a material) is the key bridge between processing and properties. Hence, a study of 
the formation and evolution of microstructures is of great interest: see, for example,~\cite{Schmitz2016} 
and references therein. Naturally, during the formation and evolution of microstructures, new interfaces may form and 
old ones might disappear; in addition, interfaces might merge or split. 

Instabilities is one of the key phenomena that leads to interesting microstructural features; 
for example, compositional instabilities in binary alloys aged inside the spinodal region of a 
miscibility gap lead to spinodal microstructures and dendritic microstructures 
result from the breaking up of planar solid-liquid interfaces during solidification. 
Martin, Doherty and Cantor, in their monograph on microstructural stability~\cite{MartinDohertyCantor} 
give a fairly comprehensive list of microstructural instabilities and classify 
them as due to chemical energy, strain energy, interfacial energy and others (such as irradiation, magnetic, 
thermal and electric fields). Our interest in this review is on microstructural instabilities that are 
influenced by elastic stresses.

Elastic stress effects on microstructures is well known -- see, for example the 
articles~\cite{FratzlPenrose,Doi,Johnson1999,VoorheesJohnson2004} or 
the monographs of Mura~\cite{Mura} and Khachaturyan~\cite{Khachaturyan}. 
Elastic stresses arise naturally during phase transformations (for example, 
lattice parameter mismatch in coherent precipitates), processing (for example, 
during the epitaxial growth of a thin film on a rigid substrate) and/or service 
and environmental conditions (for example, stressed minerals that are in contact with 
their solutions). Elastic stresses play two distinct roles in influencing 
microstructural instabilities -- namely, promotion and suppression of instabilities; 
the Asaro-Tiller-Grinfeld (ATG) instabilities is a typical example of stress induced 
instability while the suppression of spinodal decomposition is a typical example of 
stress induced suppression. 

Phase field models are ideal for the study of formation and evolution of microstructures. 
In these models, also known as diffuse interface models, the interfaces are not explicitly 
tracked. So, any topological singularity associated with the 
formation, merger, splitting and disappearance of interfaces can be handled smoothly. Further, 
interfaces are defects and hence have a positive excess free energy associated with them. 
In phase field models, this excess free energy 
associated with them can be incorporated and thus any interface related physics 
(such as Gibbs-Thomson effect, for example) 
can be automatically accounted for. Hence, phase field models have been extensively used in 
the past two decades for studying a variety of systems and their microstructures --  see~\cite{ChenAnnRevMatSci,BoettingerEtAlAnnRevMatSci,Steinbach,Voorhees,LeuvenGang,WangLi2010,
Emmerich2008,RobPhillips1998,Militzer2011} 
for some reviews.

This review is on the phase field modelling studies in elastic stress effects on microstructural instabilities. 
We will focus primarily on four elastic stress driven instabilities: 

\begin{enumerate}

\item Spinodal phase separation;

\item Particle splitting; 

\item Rafting; and, 

\item Asaro-Tiller-Grinfeld (ATG) instabilities

\end{enumerate} 

In systems that undergo elastic stress driven microstructural instabilities, 
the different constituent phase might have different moduli (that is, the system
is elastically inhomogeneous); (coherency driven) eigenstrains might be present; and, 
there might be applied traction (or imposed strains) on the system. Even though all 
these three might be present in all these four problems, for the instability to occur, 
one (or more) of these is (are) essential. For example, elastic inhomogeneity along 
with imposed strains / applied stresses is sufficient to produce ATG instabilities; rafting 
requires all three -- namely, eigenstrain, elastic inhomogeneity and applied stresses and in 
the absence of any of these it will not occur; and, suppression of spinodal and particle splitting 
can take place in the presence of eigenstrains (even if there are 
no applied stresses or imposed strains and/or elastic moduli mismatch).

There are several other microstructural instabilities, such as dendritic formation 
during solid-solid phase transformations~\cite{HenryEtAl1993,JouEtAl1997,Yoo2005}, 
buckling and wrinkling of soft films~\cite{LiEtAl2012} and liquid crystal elastomers~\cite{MbangaEtAl2010}, 
twinning~\cite{ClaytonKnap2011,LeeYoo1990,HeoEtAl2014}, 
dissolution-precipitation creep at grain boundaries in minerals~\cite{KoehnEtAl2006},
phase inversion~\cite{LeoEtAl1998,GururajanAbinandanan2007},
dynamic brittle fracture~\cite{BordenEtAl2012} and branching instability, 
cracking of surfaces~\cite{YangSrolovitz1994},
dissolution driven crack growth~\cite{StahleEtAl2007},
surface roughening instability during dynamic fracture~\cite{Gao1993}, 
step instabilities (bunching and undulation) on stressed surface~\cite{LeonardTersoff2003}, 
stress driven roughening of solid-solid interfaces~\cite{AnghelutaEtAl2008},
the destabilization of solidification and melting fronts due to stress~\cite{Colin2009},
dynamical instabilities of dislocation patterning in fatigued metals~\cite{WalgraefAifantis1985},
crystals growing on curved surfaces~\cite{MengEtAl2014},
stress induced boundary motion~\cite{GarckeEtAl2007},
martensitic transformations~\cite{ArtemevEtAl2001}, 
microstructural evolution in systems with cracks 
and voids~\cite{JinWangKhachaturyan2001,WangJinKhachaturyan2002,WangJinKhachaturyan2002JAP,JinWangKhachaturyan2003} 
and so on. In many of these instabilities elastic stresses might play an important role. However, 
in this review, we do not discuss them.

This review is organised as follows: in Section~\ref{ExperimentalResults}, we briefly describe 
some of the important and interesting experimental observations on elastic stress  effects on 
microstructural instabilities; in Section~\ref{TheoryAndModels}, 
we describe, in reasonable detail, the theoretical developments in understanding the effects of 
elastic stress on microstructural instabilities in solids. Both Section~\ref{ExperimentalResults} 
and~\ref{TheoryAndModels} are neither comprehensive nor complete; however, they are helpful in setting 
the stage for discussion of (and, in giving a perspective on) phase field modelling studies that will 
be discussed in Section~\ref{PhaseFieldModels}. In Section~\ref{Conclusions}, we conclude with a summary
and indication of future directions.

\section{Experimental observations} \label{ExperimentalResults}

The elastic stress driven microstructural instabilities are of great practical 
importance; the stress-corrosion cracking of minerals in earth's mantle, instabilities 
during the growth of thin films, formation of quantum dots, particle splitting and rafting in 
Ni-base superalloys, and suppression of spinodal decomposition are but some of the well-known 
stress driven microstructural instabilities of relevance. In this section, as noted earlier, 
we very briefly indicate some of the important experimental 
observations -- in order to set the stage for a detailed discussion of the theoretical and phase field studies.
The description here is neither complete nor comprehensive; the interested reader is referred to references~\cite{MartinDohertyCantor,FratzlPenrose,Doi,Johnson1999,VoorheesJohnson2004} for more information.

\subsection{Spinodal phase separation: suppression and promotion}
  
It is fairly well known that elastic stresses can suppress microstructural instabilities; for example,
such suppression is reported in Al-Zn~\cite{UngarEtAl1981,LoefflerEtAl1989}, Au-Pt alloys~\cite{Keijser1977}
alkali feldspars (specifically, sanidine-high albite systems)~\cite{SiplingYund1976}, 
semiconductors doped with transition metals~\cite{DietlEtAl2015}, Ci-Ni(Fe) nanolaminates~\cite{Jankowski2015}
and pyroxenes~\cite{Jantzen1984}. 

On the other hand, self-assembled quantum dots and wires in epitaxially 
grown thin films are produced using spinodal decomposition mechanism; 
see~\cite{LahiriEtAl} (and, some of the references therein~\cite{TwestenEtAl1999,LiuEtAl2000,Brunner2002,SkolnickMowbray2004,BhattacharyaEtAl2004,
RaganAtwater2005,RaganEtAl2006,BortoletoEtAl2007,AdhikaryChakrabarti2012,AquaEtAl2013}). 
However, in some of these systems, the effect of epitaxial
strain on spinodal instability is asymmetric -- for example, compressive stresses might promote
phase separation~\cite{Brunner2002,AquaEtAl2013} while tensile stresses suppress the same~\cite{MyronovEtAl2011,LiuEtAl2011}. 

\subsection{Particle splitting}

Elastic stress induced splitting instability of misfitting precipitates have been reported~\cite{MiyazakiEtAl1982,DoiEtAl1984,DoiEtAl1985,DoiMiyazaki1986,KaufmanEtAl1989,JohnsonVoorhees1992,YeomEtAl1993,
YooEtAl1995} 
in several Ni-base systems. The particle splitting instability is the opposite of coarsening; as the size becomes
larger than some critical value, the precipitate splits into doublets, quartets or octets; see~\cite{Qiu1998}, for 
an example of the wide variety of split structures that are observed.  

There are also a few studies which question the interpretation of particle splitting; 
instead it is explained as a coalescence induced~\cite{Calderon1,Calderon2,KisielowskiEtAl2007} 
or compositional heterogeneity induced (albeit in Ir-Nb system)~\cite{Harada}  microstructural feature. 
As we discuss later, there is phase field modelling based evidence supporting both the splitting and
coalescence mechanisms. 

\subsection{Rafting}

In Ni-base superalloys consisting of $\gamma^{\prime}$ precipitates (with L1$_2$, an ordered face 
centered cubic crystal structure) in nickel rich $\gamma$ (disordered fcc) matrix (as well as others with a similar
microstructure of coherent ordered precipitates in a disordered matrix), under an applied uniaxial stress, 
rafting (which is a preferential coarsening) is one of the instabilities seen: see, for example~\cite{TienCopley1971,TienCopley19712,Qiu1996,KamarajEtAl1998,MatanEtAl1999,RatelEtAl2008,TitusEtAl2012}
and the reviews of Chang and Allen~\cite{ChangAllen1991} and Kamaraj~\cite{Kamaraj2003}.
During rafting, the $\gamma^{\prime}$ precipitates coarsen preferentially under the action of the 
applied load -- either parallel or perpendicular to the direction of applied load if it is uniaxial; 
if the loading is not uniaxial, the rafting is more complex~\cite{KamarajEtAl1998}. 

Rafting leads to the destruction of an initially periodic arrangement of cuboids of precipitates 
during service and leads to a microstructure consisting of wavy precipitates with very large aspect ratios. 
Depending on the type of microstructure that coarsening leads to, it can lead to either hardening or softening 
of the microstructure: see for example~\cite{ShuiEtAl2006}. In cases of practical importance, the dislocation 
mediated plastic flow as well as twinning are known to play a crucial role in this instability: see for example~\cite{VeronEtAl1996,VeronBastie1997,ParisEtAl1997,YamashitaKakehi2006,JacquesEtAl2008,PierretEtAl2013}. 
Dislocation activity is also known to help coalesce different variants of ordered precipitates 
by helping get rid of the anti-phase boundary during rafting~\cite{ChenStobbs2004}. However, phase field 
models have indicated (as discussed below) that purely elastic stress driven (diffusional) rafting is possible.

\subsection{Asaro-Tiller-Grinfeld (ATG) and associated instabilities}

\begin{figure}[thpb]
\begin{center}
\resizebox{4in}{!}{\rotatebox{0}
{\includegraphics{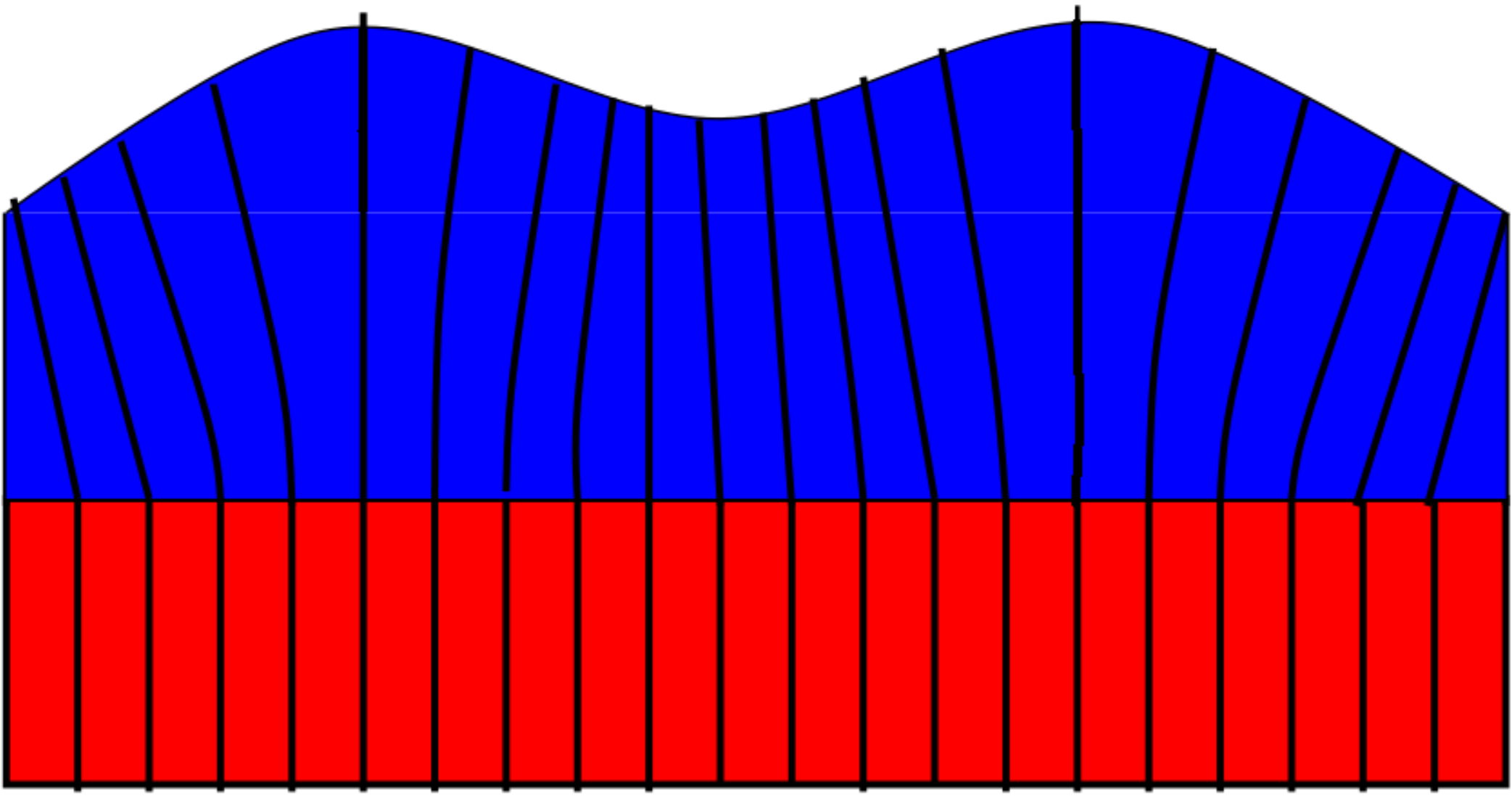}}}
\caption{Schematic explaining how the stressed film (due to the lattice parameter of the substrate (red) being
imposed on the film (blue)) can relax the stresses at the peaks. However, the stress concentration at the troughs
increases. Hence, once the undulation is set-up, it continues to grow.} 
\label{ATGSchematic}
\end{center}
\end{figure}

The surface of any non-hydrostatically stressed solid, in contact with a more compliant phase 
(be it vapour, liquid or another solid), tends to develop undulations~\cite{AsaroTiller1972,Grinfeld1984,Srolovitz1989,SridharEtAl1997}. 
This is broadly known as Asaro-Tiller-Grinfeld (ATG) instability (and will be discussed in some detail 
in the next section). ATG instabilities are reported in a wide variety of systems and conditions: for example, in
Helium IV at the solid-liquid interface~\cite{BalibarEtAl2005}, 
at the surface of SiGe films grown on Si substrates~\cite{JessonEtAl1997}, 
at the surfaces of polymeric thin films that undergo polymerisation~\cite{BerreharEtAl1992}, 
at the interfaces of minerals in contact with their solution~\cite{DenBrokMorel2001,KoehnEtAl2003,KoehnEtAl2004}, 
on the surface of pure aluminium crystals that undergo cyclic loading~\cite{KuznetsovEtAl2010,KuznetsovEtAl2012}  
and in multilayers of systems such as Si-Ge~\cite{RahmatiEtAl1996,Teichert2002} and 
Gadolinia-Silica~\cite{SahooEtAl2005}. The schematic in Fig.~\ref{ATGSchematic} (based on~\cite{JessonEtAl1997})
explains the physics behind the ATG instability: if the surface is planar, then, the imposed strains can not relax; 
however, if the surface 
develops undulations, then, the imposed strains on the film can relax at the peaks; on the other hand, the stresses
at the troughs are more than the flat surface. Hence, the undulations keep growing. In case there is growth
in the presence of such undulations, the chemical potentials are such that the atoms would preferentially attach 
to the peaks. 

These instabilities can further be classified as of two types -- namely, static and dynamic. In some cases, 
say, for example the case of minerals in contact with their solution, the ATG instability is static; that is, 
there is no growth induced movement of the mineral-solution boundary. On the other hand, in the case of 
SiGe films on Si substrates, the ATG instability is dynamic; that is, ATG is concurrent with the growth of 
the film and the instability is enhanced by the growth processes~\cite{VoorheesJohnson2004}. 
In the case of dynamic instabilities, there could be 
other elastic field mediated instabilities which are not of the ATG type: for example, Duport, Nozi\`{e}res 
and Villain~\cite{DuportEtAl1995} report on a step-bunching instability which is because of elastic 
interaction between adatoms and rest of the material during molecular beam epitaxy (and is different from ATG).
In addition, it is known that in confined systems, the dynamics of ATG instabilities could be 
different~\cite{DystheEtAl2006}. 

In the next two sections, we describe the theoretical framework and phase field models that capture several aspects
of ATG instabilities. However, the literature on ATG instabilities is too vast to be summarised here. We refer
the interested reader to the available monographs~\cite{Nozieres,PimpinelliVillain,FreundSuresh} and 
reviews~\cite{ShchukinBimberg,GaoNix,StanglEtAl,VoorheesJohnson2004,BalibarEtAl} in the literature.  

\section{Theory and models} \label{TheoryAndModels}

Solids, unlike fluids, can support non-hydrostatic stresses. Hence, the effect of such stresses on 
the thermodynamics (especially, with specific reference to phase transformations) 
is of great interest. 

As early as 1876, Gibbs~\cite{Gibbs1876} alluded to this in his 
classic work on the equilibrium of heterogeneous substances with one of the sections named as:
\begin{quote}
The conditions of internal and external equilibrium
for solids in contact with fluids with regards to all
possible states of strain of the solids
\end{quote} 
However, even four decades after Gibbs, that there was no substantial progress 
on this problem is clear from the following sentences of Bridgman (written in 1916) -- quoted from~\cite{Bridgman1916}:
\begin{quote}
The question at issue was: what is the effect on a 
transition (or melting) point of unknown extra stresses 
not hydrostatic in nature? It was a surprise to me, 
after a careful search, to find that this problem has 
received very meager attention, ...
\end{quote}
By 1950s, the contribution of surface stresses to interfacial free energy was recognised~\cite{Shuttleworth1950}.
About 45 years after Bridgman's observation, Cahn~\cite{Cahn1962} introduced the idea of coherent spinodal -- namely, 
the suppression of spinodal region due to the elastic stress effects. Larche and Cahn~\cite{LarcheCahn1985} studied the 
question of equilibrium in stressed solids with specific reference to their interaction with composition in crystalline 
solids in early 1970s; at around the same time, Asaro and Tiller~\cite{AsaroTiller1972} addressed the question of the 
equilibrium of a non-hydrostatically stressed solid in contact with its melt. 

Around 1984, Johnson and Cahn~\cite{JohnsonCahn1984} introduced the idea of elastic stress induced shape 
bifurcations. Though Asaro and Tiller addressed this question using a perturbation analysis, 
Grinfeld~\cite{Grinfeld1984}, independently, in 1986, showed that the question can be posed as 
an equilibrium problem and answered it using variational analysis; more specifically, Grinfeld showed 
that the surface of a non-hydrostatically stressed solid (however small the stress be) in contact with 
its melt will be unstable with respect to fluctuations of any wavelength (in the absence of interfacial energy) 
and that the lower wavelength limit of the fluctuations is set by the interfacial energy. 

In this section, we discuss the theoretical concepts and formulations in some detail. The rationale
behind such a detailed exposition is as follows: 

\begin{itemize}

\item The complete derivation of elastic stress induced promotion of spinodal is not
available in the literature in detail and is being presented here for the first time; 

\item There are similarities and differences between the approaches; for example, both stress induced
suppression of spinodal and ATG instability can be studied using linear stability analyses; ATG
instability itself can be studied using both perturbative variational analyses; the Eshelby
energy-momentum tensor plays a crucial role in understanding both ATG instabilities and rafting; and so on.
Having all the models described in one place helps us gain a perspective which is otherwise
missing; and, 

\item The theoretical concepts and formulations are also important from phase field modelling point of view.

{\begin{itemize}
\item The theoretical studies play a foundational role in helping us formulate the phase field models 
and checking for their correctness;

\item The theoretical studies, since they give analytical solutions under certain simplifying assumptions
and approximations, are helpful
in benchmarking the implementations of the phase field models; and,

\item The phase field models can be used to systematically relax the constraints imposed or assumptions and 
approximations made in the theoretical studies; thus, the theoretical studies are helpful in setting the
agenda for development of phase field models. Occasionally, phase field models do help in formulating
new theoretical models. 
 
\end{itemize}}

\end{itemize}

\subsection{Some basics of thermodynamics and mechanics}

As noted, starting from Gibbs, there have been continuous attempts to understand the thermodynamics of stressed solids; 
the theories are obviously based on the notions of elastic energy, the minimization of free energy, and interfacial 
energy (since, in general, during the minimization of elastic energy, it is the interfacial energy that tends to increase); 
in addition, in cases where the understanding of kinetics is attempted, the notion of chemical potential also becomes 
important~\cite{Norris1998}.

There are two broad approaches taken in the literature; one is based on the classical variational approach -- 
used by Gibbs and Grinfeld and extended by Larche and Cahn; and, the other is based on the concept of generalised forces 
(called accretive forces or configurational forces) -- used by Eshelby~\cite{Eshelby1951,Eshelby1975} and others: see for example~\cite{Gurtin1995,Barthalomeusz1995,Gurtin1996,Miranville1999,GrossEtAl2003,LiGupta2006}. In this review, we will primarily focus on 
the variational approach (though the widespread use of Eshelby energy-momentum tensor is common even in 
variational approach and is described in the next section). We refer the interested reader to the monographs of 
Maugin~\cite{Maugin1995} and Gurtin~\cite{GurtinBook} for detailed exposition of configurational force based formulations.  

Norris~\cite{Norris1998}, in a very nice exposition on the notion of chemical potential in elastically stressed solids (based on~\cite{AlexanderJohnson1985,LeoSekerka1989,Gurtin1993,Grinfeld1994,Wu1996,LarcheCahn1973,Cahn1980,LarcheCahn1985}), 
has drawn attention to several important and subtle ideas and notions that need a very careful consideration. They can be 
summarised as follows:

\begin{itemize}

\item As noted in the ATG instability case above, in all of elastic stress induced microstructural instabilities, we have to 
consider two cases: one in which diffusional processes redistribute existing material (denoted as static by us) and 
the other in which material is transferred from the surrounding (denoted as dynamic by us);

\item In continuum mechanics, usually, there are two approaches that are taken to define quantities of interest, namely, 
the Lagrangian or material coordinate and, the Eulerian or current coordinate. The expressions for chemical 
potential in these two approaches, though formally equivalent, are not symmetric and hence are to be used with care;

\item The distinction in the coordinate frames for description can also be taken further; there are two surface energy
descriptions, namely, Herring (based on Lagrangian) and Laplace (based on Eulerian); and,

\item If a variational approach is taken to the minimization of energy, then, in the case of (crystalline) solids, allowed
variations are to be defined with great care.  

\end{itemize}

The papers on the thermodynamics of stressed solids are huge in number: see for example~\cite{MullinsSekerka1985,AlexanderJohnson1986,JohnsonChiang1988,GurtinStruthers1990,Shimizu1992,
FriedGurtin1993,GurtinVoorhees1993,FriedGurtin1994,GurtinVoorhees1996,FriedGurtin1999,SekerkaCahn2004,YeremeyevEtAl2007}.
However, as seen in some of these recent attempts~\cite{Svendsen2011,Levitas2013,LevitasWarren2015} 
the quest for the formulation of thermodynamically 
consistent phase field models that obey relevant principles and laws of mechanics is far from complete.

\subsection{Eshelby energy-momentum tensor}

Eshelby introduced the notion of elastic energy-momentum tensor~\cite{Eshelby1975} defined as follows:
\begin{equation}
\bm P = W \bm 1 - \nabla \bm u^T \bm \sigma 
\end{equation}
where $W$ is the strain energy density, $\bm \sigma$ is the stress and $\bm u$ is the displacement.
The integral of the normal component of the energy-momentum tensor over a surface gives the force acting
on the defects and inhomogeneities enclosed by the surface. The introduction of such a force (called
accretive or configurational force, as noted above) acting on a defect or inhomogeneity in an elastic continuum 
is very useful; once such a force is known, the change in elastic potential due to the size and shape changes 
of the defect (as in rafting and ATG instabilities, for example) can be calculated, as we discuss below.     

\subsection{Equivalent eigenstrain and homogenisation problem}

The elastic inhomogeneity plays a key role in some of the elastic stress induced instabilities; for example, 
for both ATG instability and rafting, elastic instability is necessary; in elastically homogeneous systems, there will
be neither rafting nor ATG instability. Further, the elastic inhomogeneity can lead to surprising results as in the
case of coherent spinodal discussed below. When it comes to dealing with elastically inhomogeneous systems, there are
two approaches. In the first one, pioneered by Eshelby, the inhomogeneous problem is replaced by a homogeneous problem
by defining an equivalent eigenstrain. On the other hand, the approach taken in the composites literature is to
come up with a homogeneous effective moduli of the two phase mixture. As we see below, while the method of equivalent
eigenstrain is useful in deriving analytical solutions, the majority of phase field models that we will describe
in this review are based on the idea of homogenisation. 

\subsection{Equation of mechanical equilbrium}

As noted above, elastic inhomogeneity in systems with coherency driven eigenstrains and imposed strains or applied tractions
is the key problem in elastic stress driven microstructural instabilities. Thus, obtaining the stress and strain fields
in such systems is at the heart of the phase field models. The stress and strain fields are obtained by solving the equation
of mechanical equilibrium, namely,
\begin{equation} \label{mecheq}
\nabla \cdot \bm \sigma^{el} = 
\frac{\partial  \sigma_{ij}^{el}}{\partial r_j} 
= 0 \;\;\; {\mathrm {in}} \;\; \Omega.
\end{equation}
where $\bm \sigma^{el}$ is the elastic stress. 

Let the computational domain consists of two phases, namely, a matrix ($m$) and a precipitate ($p$).
Let us assume that both the $m$ and $p$ phases are Hookean (that is, linearly elastic):
\begin{equation}
\bm \sigma_{kl}^{el} = \bm C_{ijkl} \bm \varepsilon_{ij}^{el},
\end{equation}
where $\bm \varepsilon^{el}$ be the elastic strain and $C_{ijkl}$ is the composition (and hence, position) dependent 
elastic modulus tensor; that is, the solid is elastically inhomogeneous.

The elastic strain is derivable from the total strain $\bm \varepsilon$:
\begin{equation}
\bm \varepsilon_{ij}^{el} = \bm \varepsilon_{ij} - \bm \varepsilon_{ij}^{0}, 
\end{equation}
with $\varepsilon^{0}$ being the position dependent 
eigenstrain (misfit strain) tensor field.
The total strain $\bm \varepsilon_{ij}$ is compatible; that is, it is derivable from
the displacement field $\bm u$ as follows:
\begin{equation}
\bm \varepsilon_{ij} =\frac{1}{2} (\bm \nabla \bm u + \bm \nabla \bm u^T) = 
\frac{1}{2} \left\{ \frac{\partial u_{i}}{\partial
r_{j}} + \frac{\partial  u_j}{\partial r_{i}}\right\}.
\end{equation}

Using the symmetry properties of the moduli tensor, the equation of mechanical equilibrium can be written as:
\begin{equation} \label{EqMechEq}
\frac{\partial} {\partial r_j} \left( C_{ijkl} \frac{\partial u_i}{\partial r_k} - \varepsilon^{0}_{jl}\right)
= 0 \;\;\; {\mathrm {in}} \;\; \Omega.
\end{equation}

In this partial differential equation, the coefficients are composition (and hence position) dependent. 
Such partial differential equations with varying coefficients require the technique of homogenisation 
for their solution~\cite{NematNasserHori,Torquato}.
In the next section, we discuss the homogenisation technique.

\subsection{Homogenisation} \label{homogenisation}

Let us assume the following composition dependence:
\begin{equation} \label{eigenstrain}
\varepsilon^{0}_{ij} (c) = \beta(c) \varepsilon^{T} \delta_{ij},
\end{equation}
where, $\varepsilon^{T}$ is a constant that determines the strength of
the eigenstrain, $\delta_{ij}$ is the
Kronecker delta, and $\beta(c)$
is a scalar function of composition; and,
\begin{equation} \label{elastmod}
C_{ijkl} (c) = C_{ijkl}^{\mathrm {eff}} 
+ \alpha(c) \Delta C_{ijkl},
\end{equation}
where $\alpha(c)$ is a scalar
function of composition, 
and,
\begin{equation} \label{deltaC}
\Delta C_{ijkl} = C_{ijkl}^{p} - C_{ijkl}^{m}
\end{equation}
where, $C_{ijkl}^{p}$ and $C_{ijkl}^{m}$ are the elastic moduli tensor
of the $p$ and $m$ phases respectively, and $C_{ijkl}^{\mathrm {eff}}$ is an
``effective" modulus. 

With these composition dependence for the eigenstrains and elastic moduli, we want to solve the Eq.~\ref{EqMechEq}. 
The computational domain is assumed to be a representative volume element; that is, we will assume the composition 
field to be periodic on the domain. This implies that some of the fields that are derived from composition
such as eigenstrains and elastic moduli are also periodic; on the other hand, applied tractions will have to 
be anti-periodic. Thus, Eq.~\ref{EqMechEq} has to be solved with such periodic and anti-periodic boundary conditions. 
In addition, there are also boundary conditions of either applied
traction or imposed strains. The imposition of these boundary conditions is achieved using 
`homogenisation': that is, we define the mean strain and stress in the computational domain 
as follows:
\begin{equation} \label{ave}
\langle \{ \varepsilon_{ij}\} \rangle = E_{ij},
\end{equation}
where $\bm \varepsilon$ is the total strain, and 
the symbol $\{ \cdot \}$ is defined as follows:
\begin{equation} \label{avedefn}
\langle \{\cdot\} \rangle  = \frac{1}{V} \int_{\Omega} \{\cdot\} d\Omega,
\end{equation}
where $V$ is the volume of the representative domain $\Omega$; and,
\begin{equation}
\langle \{\sigma^{el}_{ij} \} \rangle 
= 
\frac{1}{V} \int_{\Omega} \sigma^{el}_{ij} d\Omega.
\end{equation}
The mean stress thus calculated should equal the applied stress
$\bm \sigma^{A}$~\cite{Anthoine1995,Michel1999,JinWangKhachaturyan2001}. This conclusion 
namely, that the mean stress should equal the applied stress, is arrived
at using homogenisation assumption by some authors~\cite{Anthoine1995,Michel1999}
while Jin et al~\cite{JinWangKhachaturyan2001} used a variational 
approach. 

The Eq.~\eqref{elastmod} is written assuming that in spite of the inhomogeneities
at the microscopic scale, the domain $\Omega$ behaves as if it is a single homogeneous block
with an ``effective" elastic modulus $C_{ijkl}^{\mathrm {eff}}$;
the local microscopic perturbations in the elastic moduli (with respect to 
$C_{ijkl}^{\mathrm {eff}}$) are described using the
difference between the elastic constants of the $p$ and $m$ phases ($\Delta C_{ijkl}$).
As noted, the relevant boundary conditions are imposed strains or applied traction;
since we will be using a spectral technique, we assume that the domain is periodic;
hence, if macroscopic system is subjected to a 
homogeneous stress state $\sigma^{A}$, then, the applied traction 
on the boundaries of the domain $\Omega$ will be anti-periodic; i.e., $\bm \sigma
\cdot {\mathbf n}$, 
is opposite on opposite sides of 
$\partial \Omega$ with 
$\mathbf n$ being the unit normal to the 
boundary~\cite{Michel1999,Anthoine1995}.

The definition of periodic strain is again a result of homogenisation.
By the imposed periodic boundary condition, the solution to the equilibrium
equation (Eqn.~\eqref{mecheq}) will be such that the strain field 
$\varepsilon ({\mathbf r})$ is periodic on $\Omega$. However, we have posed
the equation of mechanical equilibrium in terms of the displacement field. 
Since the strains are derived from displacements by differentiation, the 
displacement field ${\mathbf u} ({\mathbf r})$ which 
gives rise to such periodic strain fields can always be written as follows~\cite{Anthoine1995}:
\begin{equation} \label{disp}
{\mathbf u} = {\mathbf E} \cdot {\mathbf r} + {\mathbf u^{\star}},
\end{equation}
where, ${\mathbf u^{\star}}$ is a displacement field that is periodic
on $\Omega$ and ${\mathbf E}$ is a constant, homogeneous strain tensor. 
$\bm E$ can be assumed to be symmetric (without loss of generality) since 
the antisymmetric part corresponds to a rigid rotation of the cell.
The `homogenisation' implies~\cite{Anthoine1995} that ${\mathbf E}$ 
is the mean strain tensor of the cell (see Appendix D in~\cite{GuruThesis}).

Let $\varepsilon^{\star}$ be the
periodic strain; then, the strain we derive from the displacement equation~\eqref{disp}
becomes (see Appendix D in~\cite{GuruThesis}),
\begin{equation}
\varepsilon_{ij} 
= 
E_{ij} + \varepsilon^{\star}_{ij},
\end{equation}
where, 
\begin{equation}
\varepsilon^{\star}_{ij} 
= \frac{1}{2} \left\{ \frac{\partial u^{\star}_{i}}{\partial
r_{j}} + \frac{\partial u^{\star}_j}{\partial r_{i}}\right\},
\end{equation}
and the equation of mechanical equilibrium~\eqref{mecheq} is
\begin{equation}
\frac{\partial}{\partial r_j}
\{ 
C_{ijkl} 
( E_{kl} +\varepsilon^{\star}_{kl} - \varepsilon^{0}_{kl} ) 
\}
= 0.
\end{equation}

Using the mean stress equation, it is easy to show that
\begin{equation} \label{homstrain}
E_{ij} = S_{ijkl} \; (\sigma^{A}_{kl} + \langle \{\sigma^{0}_{kl} \}
\rangle
- \langle \{ \sigma^{\star}_{kl} \} \rangle).
\end{equation}
where,
\begin{equation} \label{Sdef}
S_{ijkl} = (\langle \{C_{ijkl}\} \rangle)^{-1},  \;
\langle \{\sigma_{ij}^{\star} \} \rangle = 
\langle \{C_{ijkl} \varepsilon_{kl}^{\star}\} \rangle, \; 
{\mathrm {and}} \;
\langle \{\sigma_{ij}^{0} \} \rangle 
=  \langle \{C_{ijkl} \varepsilon_{kl}^{0} \} \rangle.
\end{equation}
and, $\bm \varepsilon^0$ is the composition (and hence) dependent eigenstrain
and $\bm \varepsilon^{\star}$ is the periodic strain.

So, we obtain
\begin{equation} \label{appstress}
\sigma_{ij}^{A} = 
\frac{1}{V} \int_{\Omega} 
C_{ijkl} 
( E_{kl} +\varepsilon^{\star}_{kl} - \varepsilon^{0}_{kl} ) d\Omega.
\end{equation}

Thus, using homogenisation, the equation of mechanical equilibrium can be restated as follows:
\begin{center}
\parbox{4in}{
Given a periodic composition field $c$ on $\Omega$,\\
solve the {\bf equation of mechanical equilibrium} \\
\begin{equation} \label{eqeq}
\frac{\partial}{\partial r_j}
\{ 
C_{ijkl} 
( E_{kl} +\varepsilon^{\star}_{kl} - \varepsilon^{0}_{kl} ) 
\}
= 0 \; {\mathrm {on}} \; \Omega,
\end{equation}
with the {\bf constraint}\\
\begin{equation} \label{constraint}
E_{ij} = S_{ijkl} (\sigma^{A}_{kl} + \langle \{\sigma^{0}_{kl} \}
\rangle
-  \langle \{\sigma^{\star}_{kl}\} \rangle)
\end{equation}
and the {\bf boundary condition}\\
\begin{equation}
\varepsilon^{\star}_{kl} \; {\mathrm {is}} \; {\mathrm {periodic}} \; 
{\mathrm {on}} \; \Omega.
\end{equation}
}
\end{center}
In this formulation, now it is easy to implement
an overall prescribed strain ($E_{ij} \neq 0$). 
It is also possible to prescribe overall stress using the
same quantity; this approach of stress control is known as
``stress-control based on strain-control" and is described
in~\cite{Michel1999}.

Substituting for $C_{ijkl}$, and $\varepsilon^{0}_{kl}$ in terms of
composition, and 
$\varepsilon^{\star}_{kl}$ in terms of the displacement field in
Eqn.~\eqref{eqeq}, and
using the symmetry properties of the elastic constants and strains, 
we obtain
\begin{eqnarray}
\frac{\partial}{\partial r_j} \left\{ 
[C_{ijkl}^{\mathrm {eff}} + \alpha(c) \Delta C_{ijkl}]
\left (E_{kl} + \frac{\partial u^{\star}_{l}({\mathbf r})}{\partial r_{k}} 
- \varepsilon^{T} \delta_{kl} \beta(c) \right) \right\}  &&\\ \nonumber
&=& 0. 
\end{eqnarray}
\begin{eqnarray} \label{eq_mech_eq_final}
\left[ C_{ijkl}^{\mathrm {eff}}\frac{\partial^{2}}{\partial{r_{j}}\partial{r_{k}}} 
+ \Delta C_{ijkl} \frac{\partial}{\partial{r_{j}}} \left( \alpha(c) 
\frac{\partial}{\partial{r_{k}}} \right) \right]
u^{\star}_{l}({\mathbf r}) 
&=& C_{ijkl}^{\mathrm {eff}} 
\varepsilon^{T} \delta_{kl} \frac{\partial \beta(c)}{\partial r_{j}} \\
\nonumber
&& - \Delta C_{ijkl} E_{kl}
\frac{\partial \alpha(c)}{\partial r_{j}} \\ \nonumber
&& + \Delta C_{ijkl} \varepsilon^{T} \delta_{kl}
\frac{\partial \{\alpha(c)\beta(c)\}}{\partial r_{j}}.
\end{eqnarray}

\subsection{Fourier transform based iterative solution to the equation of 
mechanical equilibrium}\label{Eq_mech_eq_soln}

The equation of mechanical equilibrium is typically solved using finite element technique. 
However, the finite element method requires
meshing of the domain with denser mesh close to the interfaces. In phase field models, 
this cost of meshing can be too high; 
for example, in a system undergoing spinodal decomposition, the entire domain, at least in the 
early stages of decomposition, consists only 
of interfaces (albeit at various stages of formation). 
In addition, as the microstructure evolves the interfaces continuously merger 
and split, and new interfaces appear while old ones disappear.
Hence, in the phase field literature, an alternate iterative method based on spectral 
techniques is widely used for solving 
the equation of mechanical equilibrium; in this section, we describe the method.  
The disadvantage with this method, is, of 
course that it is iterative (though there are methods proposed, based on FFT to 
tackle these situations also: see~\cite{YuEtAl2005}). So, when the 
`contrast' (that is, the ratio of elastic moduli of the two phases) 
is too high, the iterations
take much longer to converge making finite element implementations 
(which solve the problem in one step) competitive.  The spectral 
techniques are based on Fourier transform; hence the use of numerically efficient Fast Fourier 
Transform codes (such as FFTW~\cite{FFTW}) is widespread in the implementations of this method.

The iterative technique for solving the equation of mechanical equilibrium using Fourier transforms is well known~\cite{KhachSemTsakalakos,Anthoine1995,WangKhachaturyan1995,Michel1999,LiChen1999,JinWangKhachaturyan2001,
HuChen2001,GuruAbiRafting2007}: our description below is based on~\cite{GuruThesis}. Since we 
are using the Fourier transform based technique, in the computational domain $\Omega$
the fields are either periodic (for example, the composition, and the moduli and eigenstrains that follow the composition) 
or anti-periodic (for example, the applied traction). As noted in the previous section, the assumption of periodicity of 
computational domain is also justified physically since the domain is the representative volume element.

\subsubsection{Zeroth order approximation}

Assume $\Delta C_{ijkl} = 0$; the equation of mechanical equilibrium~\eqref{mecheq} simplifies to
\begin{equation}
C_{ijkl}^{\mathrm {eff}}\frac{\partial^{2}u^{\star}_{l}({\mathbf r})}
{\partial{r_{j}}\partial{r_{k}}} 
= C_{ijkl}^{\mathrm {eff}} \varepsilon^{T} \delta_{kl} 
\frac{\partial \beta(c)}{\partial r_{j}}.
\end{equation}

Let $ \sigma_{ij}^{T} = C_{ijkl}^{\mathrm {eff}} \varepsilon^{T}
\delta_{kl}$:
\begin{equation} \label{zeroth-approx}
C_{ijkl}^{\mathrm {eff}}\frac{\partial^{2}u^{\star}_{l}({\mathbf r})}
{\partial{r_{j}}\partial{r_{k}}} 
= \sigma_{ij}^{T} \frac{\partial \beta(c)}{\partial r_{j}}.
\end{equation}

Let $G^{-1}_{il}$ as $C_{ijkl} g_{j} g_{k}$ 
(where ${\mathbf g}$ is
the vector in the Fourier space). Then the solution (in the Fourier space)
for the equation above is~\cite{GuruThesis}
\begin{equation} \label{solution_one}
\left\{(u^{\star}_{l})^{0}\right\}_{{\mathbf g}} 
= -J G_{il} \sigma_{ij}^{T} g_{j} 
\{\beta(c)\}_{{\mathbf g}},
\end{equation}
where the superscript on $u^{\star}_{k}$ denotes the order of
approximation, and $J$ is $\sqrt{(-1)}$.

\subsubsection{Higher order approximations}

The zeroth order approximation can be refined to obtain the first order solution. 
This process can be continued to higher orders; knowing the $(n-1)$th order solution,  
the $n$th order refined solution as follows:
\begin{equation} \label{solution_two}
\{(u^{\star}_{l})^{n}\}_{{\mathbf g}} = -J G_{il} \Lambda_{ij}^{n-1} g_{j},
\end{equation}
where
\begin{eqnarray}
\Lambda_{ij}^{n-1} & = & \sigma^{T}_{ij} \{\beta(c)\}_{{\mathbf g}} 
- \Delta C_{ijmn} E^{n-1}_{mn} \{\alpha(c)\}_{{\mathbf g}} \\ \nonumber
&& + \Delta C_{ijmn} \varepsilon^{T} \delta_{mn} 
\{\alpha[c({\mathbf r})] \; \beta[c({\mathbf r})] \}_{{\mathbf g}}  
- \Delta C_{ijmn} 
\left\{ \alpha[c({\mathbf r})] \frac{\partial
(u^{\star}_{m})^{n-1}({\mathbf r})}
{\partial r_{n}} \right\}_{{\mathbf g}} 
\end{eqnarray}

\subsection{Spinodal phase separation: suppression and promotion}

The elastic field induced suppression of spinodal decomposition is very well known~\cite{HilliardPhaseTransArticle}. However, that elastic strains can promote spinodal decomposition is not widely recognised~\cite{LahiriEtAl}. 
In this section, we describe the analyses of both these scenarios. Unlike the other 
theoretical studies described in this section (which are sharp interface models), the description of
spinodal decomposition necessarily involves building a phase field model. So, we describe the classical
(sharp interface model of) diffusion before discussing the models of spinodal decomposition.

\subsubsection{Classical diffusion equation and its failure}

Let us consider the classical diffusion equation: it is based on the constitutive law 
(known as Fick's first law) which connects the atomic flux (denoted by $\mathbf{J}$) to 
concentration gradient ($\nabla c$) through the material property known as diffusivity tensor ($\mathbf{D}$):
\begin{equation}
\mathbf{J} = - \mathbf{D} \nabla c
\end{equation}
The diffusivity is a second rank tensor; hence, in isotropic and cubic systems it is replaced by $D \delta_{ij}$ where $\delta_{ij}$ is the
Kronecker delta and $D$ is a material constant known as diffusion coefficient. For the rest of this review, 
we use diffusion coefficient.  

Using the law of conservation of mass in differential form,
\begin{equation}
\frac{\partial c}{\partial t} = - \nabla \cdot J
\end{equation}
along with Fick's first law, one obtains the classical diffusion equation (which is also called Fick's second law):
\begin{equation}
\frac{\partial c}{\partial t} = \nabla \cdot D \nabla c
\end{equation}

If the diffusivity is assumed to be a constant (that is, not a function of composition, and hence, position), we obtain
\begin{equation}
\frac{\partial c}{\partial t} = D \nabla^2 c
\end{equation}

This equation indicates that the rate of change of composition at any point is given by the curvature of 
the composition profile at that point; hence, if one assumes a sinusoidal composition profile, one can see that the compositional heterogeneities will be evened out with time. However,
in some systems, it was known that compositional heterogeneities grow with time 
(leading to phase separation -- called spinodal decomposition) instead of getting evened out, 
giving rise to the so-called ``up-hill" diffusion. 

One way that the compositional heterogeneities will grow is if the diffusivity is a negative constant. 
The negative value of diffusion
coefficient can be explained if the Fick's first law is modified using the knowledge of classical thermodynamics, 
namely, that it is the 
chemical potential gradients that drive diffusion and not compositional heterogeneities. In other words, 
the modified Fick's first law
states that
\begin{equation}
\mathbf{J} = - {\bm M} \nabla \mu
\end{equation}
where $\bm M$ is the mobility tensor and $\mu$ is the chemical potential, defined as $\frac{\partial (G/N_v)}{\partial c}$ where $G/N_V$ is the Gibbs free energy per atom ($N_V$ is the number of atoms per mole, and $G$ is the Gibbs free energy per mole). Here again, in isotropic and cubic systems $\bm M$ can be replaced by $M \delta_{ij}$ where $M$ is the mobility;
for the rest of this review, we use $M$ and not the mobility tensor. Note that in condensed systems, 
the Gibbs and Helmholtz free energies can be assumed to be the same. 

Combining this constitutive law with the law of conservation of mass, one obtains
\begin{equation}
\frac{\partial c}{\partial t} = \nabla M \nabla \mu
\end{equation}
If we assume the mobility to be a constant independent of composition, one can see that the mobility and diffusivity are related through
the relationship:
\begin{equation}
D = M \nabla^2 (G/N_V)
\end{equation}
In other words, when the sign of the curvature of the free energy versus composition curve is negative, one expects the diffusivity to become negative and the up-hill diffusion to take place and the classical diffusion takes place when the curvature of the free energy versus composition curve is positive. Thus, the point where the curvature of the free energy versus composition curve becomes zero, (that is, 
$G^{\prime \prime} = \frac{\partial^2 G}{\partial c^2} = 0$) defines the region in which spinodal decomposition
will take place. 

\subsubsection{Going beyond classical diffusion equation}

The chemical potential based constitutive law can (partially) explain spinodal phase separation -- namely,
it can explain the `up-hill' diffusion. However, it does not explain all the observed phenomena. Specifically, 
if the diffusivity is negative, one expects sinusoidal composition profiles of any wavelength to 
grow with time; further, smaller the wavelength, 
faster will be the growth of such composition waves (since shorter diffusion 
distances lead to smaller diffusion times).
However, in such up-hill diffusion cases, it was known that only compositional heterogeneities
 with a wavelength greater than certain critical wavelength grow~\cite{HilliardPhaseTransArticle}; 
and the amplitude of any wavelengths shorter than the critical wavelength diminishes with time. 

Cahn~\cite{Cahn1961} showed that the critical wavelength is a consequence of the (incipient) interface energy (due
to the formation of the two phases). As we discuss in the next section, incorporation of this
interfacial contribution through a (positive) constant known as the gradient energy coefficient ($K$)
leads to a modified diffusion equation (in 1-D):
\begin{equation} 
\frac{\partial c}{\partial t} =  \left(\frac{M}{N_V} \right) \left[ G^{\prime \prime} \frac{\partial^2 c}{\partial x^2} 
- 2 K \frac{\partial^4 c}{\partial x^4}\right]  
\end{equation}

Further, in solids, if the two phases are coherent, the phase separation can also lead to eigenstrains. Let $\eta$
be the strength of the eigenstrain ($\varepsilon^{T}$), where $\eta$ is the Vegard's coefficient:
\begin{eqnarray}
\eta=\frac{1}{a_{0}} \frac{da}{dc}\Big{|}_{c=c_{0}},
\label{eq_phase_field:eight}
\end{eqnarray}
where, $c_0$ is the overall alloy composition, 
$a$ and $a_{0}$ are the composition dependent lattice parameter and the lattice parameter 
of the reference (i.e., of the homogeneous alloy) respectively. In such elastically stressed systems,
Cahn~\cite{Cahn1961,Cahn1962,Cahn1965,Cahn1967,Tiapkin1977} showed that elastic strains can suppress spinodal decomposition, 
leading to the description of what is known as coherent spinodal, which has now become standard textbook material~\cite{PorterEasterling}. 

To understand coherent spinodal, let us consider the modified diffusion equation of Cahn in 1-D (including elastic strain)~\cite{HilliardPhaseTransArticle}:
\begin{equation} \label{ModDiffEq}
\frac{\partial c}{\partial t} =  \left(\frac{M}{N_V} \right) \left[ (G^{\prime \prime} + 2 \eta^2 Y) \frac{d^2 c}{dx^2} 
- 2 K \frac{d^4 c}{dx^4}\right] . 
\end{equation}

Consider a spatial composition profile described by
\begin{equation}
c - c_0 = \int A(\beta) \exp{(\mathrm{i} \beta x)} d\beta
\end{equation}
where $c_0$ is the overall alloy composition and $\beta$ is the wavenumber (related to the wavelength $\lambda$ as $2\pi/\lambda$).
If we substitute this profile in Eq.~\ref{ModDiffEq},
we obtain the differential equation:
\begin{equation}
\frac{dA}{dt} = -\left(\frac{M}{N_V} \right) \left[G^{\prime \prime} + 2 \eta^2 Y + 2 K \beta^2 \right] A \beta^2
\end{equation}
The solution of this differential equation is:
\begin{equation}
A(\beta, t) = A(\beta,0) \exp{(R(\beta)t)}
\end{equation}
where $R(\beta) = -\left(\frac{M}{N_V} \right) \left[G^{\prime \prime}  + 2 \eta^2 Y+ 2 K \beta^2 \right] \beta^2$.
From this solution, it is clear that the sign of $R(\beta)$ determines whether a given composition profile will
grow or not. 

Let us first consider the case where there is no eigenstrain ($\eta = 0$). If $G^{\prime \prime} > 0$, the composition fluctuations die out irrespective of $\beta$. However, if $G^{\prime \prime} < 0$, there is a critical wavenumber 
$\beta_c = \sqrt{\frac{- G^{\prime \prime}}{2 K}}$ (obtained by equating 
$\left[G^{\prime \prime}  + 2 K \beta^2 \right]$ to zero). Any wavenumber smaller than this will grow and any 
wavenumber larger than this will grow. The point at which $G^{\prime \prime} = 0$ is known as the (chemical) spinodal.
Let us now assume $\eta \neq 0$. In this case, the critical wavenumber is given by 
$\beta_c = \sqrt{\frac{- G^{\prime \prime} - 2 \eta^2 Y}{2 K}}$. Hence, the point at which 
$G^{\prime \prime} + 2 \eta^2 Y = 0$ is known as coherent spinodal. 
Since $Y \eta^2$ is a positive quantity, it is clear that the elastic stresses suppress spinodal decomposition.

\subsection{Stability analysis in an elastically inhomogeneous system under imposed strains}

The iterative procedure described above (in Sec.~\ref{Eq_mech_eq_soln}) can  be used to
obtain (approximate) analytical solutions in certain elastically inhomogeneous systems; 
for example, in a system with a sinusoidal composition profile under plane stress approximation. 
Obtaining such an analytical solution helps us 
extend the spinodal analysis of Cahn. 

Let us assume the following composition dependence for the $\alpha$ and $\gamma$ functions, 
namely, the composition dependence of the elastic moduli and 
eigenstrain (described in Sec.~\ref{homogenisation}), respectively:
\begin{equation}
\alpha(c) = \gamma(c) = c - c_{0}
\end{equation}
Note that in this part of the derivation we have used $\gamma$ for the composition dependence of
eigenstrain (instead of $\beta$ as earlier) to avoid confusion with the wavenumber denoted by $\beta$.

\subsubsection{\label{Zeroth Order} Zeroth order approximation}

The solution to Eq.~(\ref{EqMechEq}) assuming a homogeneous modulus, in Fourier space is given by~\cite{GuruAbiRafting2007}
(as shown in Eq.~\ref{Zeroth Order}):
\begin{eqnarray}
{\{{(u_{l}^{*})}^{0}}\}_{\bm{g}}=-\emph{J} G_{li} \sigma_{ij}^{T} {\{\gamma(c)\}}_{\bm{g}} g_{j}, 
\label{eq_zeroth_order:one}
\end{eqnarray}
where, $G_{il}^{-1}=C_{ijkl}^{eff} g_{j} g_{k}$ and $\sigma_{ij}^{T}=C_{ijkl}^{eff} \eta \delta_{kl}$; ${\{ \cdot \}}_{\bm{g}}$ denotes the quantity inside brackets to be in Fourier space; $g_{j}$ denotes the $j$th component of the Fourier space vector $\bm{g}$.

By adopting a similar approach to Cahn's\cite{Cahn1961}, we assume, $c-c_{0}=A \cos \beta x$, where ${2\pi/\beta}$ represents any generic wavelength. So, in Fourier space, we get:
\begin{eqnarray}
{\left[ \begin{array}{c} {(u_{1}^{*})}^{0} \\ {(u_{2}^{*})}^{0} \end{array} \right]}_{\bm{g}}=\left[ \begin{array}{c} \frac{\eta (1+\nu)}{\beta} A\pi J [\delta(\bm{g}+\bm{\beta})-\delta(\bm{g}-\bm{\beta})] \\ 0 \end{array} \right],
\label{eq_zeroth_order:two}
\end{eqnarray}
which when reverted back to the real space, we get:
\begin{eqnarray}
{\left[ \begin{array}{c} {(u_{1}^{*})}^{0} \\ {(u_{2}^{*})}^{0} \end{array} \right]}_{\bm{r}}=\left[ \begin{array}{c} \frac{\eta (1+\nu)}{\beta} A sin\beta x \\ 0 \end{array} \right].
\label{eq_zeroth_order:three}
\end{eqnarray}
Thus the only non-zero periodic strain component is:
\begin{eqnarray}
{(\epsilon_{11}^{*})}^{0}=\eta (1+\nu) A cos\beta x=
\eta(1+\nu)(c-c_{0}).
\label{eq_zeroth_order:four}
\end{eqnarray}

\subsubsection{\label{First Order} First order approximation}

As we are interested to derive expressions valid for very early stages, we can neglect the non-linear terms in the expression given in Ref.~\cite{GuruAbiRafting2007}. Thus the expression for the periodic displacement field becomes:
\begin{eqnarray}
{\{{(u_{l}^{*})}^{1}}\}_{\textbf{g}}=-\emph{J} G_{li} \sigma_{ij}^{T} \{\gamma(c)\}_{\textbf{g}}g_{j} +\nonumber \\
\emph{J} G_{li}\bigtriangleup C_{ijmn} E_{mn}^{0} \{ \alpha(c)\}_{\textbf{g}}  g_{j},
\label{first_order_eq:one}
\end{eqnarray} 
where, $E_{mn}=e\delta_{mn}$, with the reference being the unstrained homogeneous alloy lattice. The first term in the right
hand side (RHS) of Eq.~(\ref{first_order_eq:one}) is the solution to the zeroth order approximation which we already have in Eq.~(\ref{eq_zeroth_order:two}). So, we are going to consider only the second term in the RHS of Eq.~(\ref{first_order_eq:one}). Denoting it by $\bm{v}$, we get:
\begin{eqnarray}
{\left[ \begin{array}{c} v_{1}^{*} \\ v_{2}^{*} \end{array} \right]}_{\bm{g}}=\left[ \begin{array}{c} -\frac{ye (1+\nu)}{\beta} A\pi J [\delta(\bm{g}+\bm{\beta})-\delta(\bm{g}-\bm{\beta}) \\ 0 \end{array} \right],
\label{first_order_eq:two}
\end{eqnarray}
where $y$ is given as:
\begin{eqnarray}
y=\frac{1}{Y_{0}}\frac{dY}{dc}\Big{|}_{c_{0}}=\frac{\Delta Y}{Y_{0}},
\label{first_order_eq:three}
\end{eqnarray}
with $Y_{0}$ denoting the Young's modulus of the homogeneous alloy and $\Delta Y=Y^{p}-Y^{m}$. In the real space:
\begin{eqnarray}
{\left[ \begin{array}{c} v_{1}^{*} \\ v_{2}^{*} \end{array} \right]}_{\bm{r}}=\left[ \begin{array}{c} -\frac{ye(1+\nu)}{\beta} A sin\beta x \\ 0 \end{array} \right].
\label{first_order_eq:four}
\end{eqnarray}
So the periodic diplacement field obtained from the First order approximation is given by:
\begin{eqnarray}
{\left[ \begin{array}{c} {(u_{1}^{*})}^{1} \\ {(u_{2}^{*})}^{1} \end{array} \right]}_{\textbf{r}}={\left[ \begin{array}{c} {(u_{1}^{*})}^{0}+v_{1}^{*} \\ {(u_{2}^{*})}^{0}+v_{2}^{*} \end{array} \right]}_{\textbf{r}}=\left[ \begin{array}{c} \frac{(\eta-ye)(1+\nu)}{\beta} A sin\beta x \\ 0 \end{array} \right].\nonumber \\
\label{first_order_eq:five}
\end{eqnarray}
The periodic strain field is given by,
\begin{eqnarray}
{(\epsilon_{11}^{*})}^{1}=(\eta-ye)(1+\nu) A cos\beta x=
(\eta-ye)(1+\nu)(c-c_{0}),\nonumber \\
\label{first_order_eq:six}
\end{eqnarray}
with the other strain components being zero.

\subsubsection{\label{Elastic_enrgy}Elastic Energy}
The total strain energy is given by:
\begin{eqnarray}
F_{el}=\frac{1}{2}\int_\Omega \sigma_{ij}^{el} \epsilon_{ij}^{el} d\Omega ,
\end{eqnarray}
Substituting for the elastic modulus tensor components in terms of $Y$ and $\nu$ and setting  $\epsilon_{11}^{*}=P(c-c_{0})$ where $P=\eta(1+\nu)$  for zeroth order approximation and $P=(\eta-ye)(1+\nu)$ for first order approximation respectively, we get :
\begin{eqnarray}
W_{E}=\frac{1}{2}\int_{\Omega}\Big[\frac{Y_{0}}{1-\nu^{2}}P^{2}-\frac{2Y_{0}(1+\nu)}{1-\nu^{2}}\eta P+\nonumber \\ 
\frac{2Y_{0}(1+\nu)}{1-\nu^{2}}\eta^{2}+ \frac{2\Delta Y (1+\nu)}{1-\nu^{2}}eP-\frac{4\Delta Y (1+\nu)}{1-\nu^{2}} e \eta \Big]{(c-c_{0})}^{2} d\Omega.\nonumber \\
\label{eq_elastic_energy:one}
\end{eqnarray}
where we have neglected the constant terms as they do not contribute to the elastic chemical potential. Proceeding as in Ref.~\cite{Cahn1961}, we get the expression for the maximally growing wave-number for the zeroth order approximation as:
\begin{eqnarray}
\beta_{max}={\left[- \frac{\left( \frac{\partial ^{2} f_{0}}{\partial c^{2}} + \eta^{2}\frac{Y_{0}}{N_{V}} \left(1-\frac{2ey}{\eta} \right) \right)}{4 K}  \right]}^{\frac{1}{2}},
\label{eq_elastic_energy:two}
\end{eqnarray}
while for the first order approximation, we get:
\begin{eqnarray}
\beta_{max}={\left[- \frac{\left( \frac{\partial ^{2} f_{0}}{\partial c^{2}} + \eta^{2}\frac{Y_{0}}{N_{V}} \left(1-\frac{2ey}{\eta} \right) -\frac{Y_{0}{(1+\nu)}^{2} y^{2}e^{2}}{N_{V}(1-\nu^{2})} \right)}{4 K}  \right]}^{\frac{1}{2}}. \nonumber \\
\label{eq_elastic_energy:three}
\end{eqnarray}

The zeroth order approximation given by Eq.~(\ref{eq_elastic_energy:two}) has the following salient features:
\begin{enumerate}
\item On setting $\eta=0$, we recover Cahn's results\cite{Cahn1961} for a system  with no elastic misfit. 

\item Cahn's expression of the maximally growing wavenumber\cite{Cahn1961} for a system with a non-zero $\eta$ (but the elastic modulus tensor being a constant) is recovered by setting either $y=0$ or $e=0$. Thus, there is no influence of a non-zero $y$ when $e=0$, and vice versa.

\item When $2ey/ \eta>0$, the positive contribution from the elastic energy goes down, which manifests as larger maximally growing wavenumbers (i.e., shorter wavelengths) compared to that predicted by Cahn's theory for a system with homogeneous modulus.
\end{enumerate} 

Compared to Eq.~(\ref{eq_elastic_energy:two}) Eq.~(\ref{eq_elastic_energy:three}) has an additional term. 
This new term (let this be called \emph{A}) is:
\begin{eqnarray}
A=-\frac{Y_{0}{(1+\nu)}^{2} y^{2}e^{2}}{N_{V}(1-\nu^{2})}.
\label{results_first_order:one}
\end{eqnarray} 
This new term has the following features:
\begin{enumerate}
\item There is no $\eta$ in Eq.~(\ref{results_first_order:one}). So its contribution is independent of the value of the misfit in the system.

\item The energy contribution is negative regardless of the signs of $e$ or $y$. Thus, for given values of 
$e$ and $y$, from the first order approximation we get a $\beta_{max}$ which is larger than that obtained 
from the zeroth order approximation. In other words, this term promotes phase separation. 
\end{enumerate}

We will use these expressions in the next section to show how the imposed strains in these systems can promote
spinodal decomposition even outside of chemical spinodal.

\subsection{Particle splitting}

The particle splitting instability is attributed to elastic interaction energy~\cite{DoiEtAl1985,JohnsonVoorhees1987,
KhachaturyanEtAl1988,Doi1992,JohnsonVoorhees1992} between the misfitting precipitates; that is, for a given volume, 
if there are more than one precipitate aligned along certain directions of the matrix, the interaction energy 
between such misfitting  precipitates is predicted to lead to a reduction in energy which more than compensates 
for the increase in interfacial energy during splitting. However, this is not the only explanation. As we discuss 
in the next section, phase field models have shown that nucleation at dislocations, anti-phase domains of ordered precipitates, growth instabilities, particle coalescence, and, applied stress can also lead to split patterns. 
Thus, currently, the theoretical analysis of this instability is neither complete nor comprehensive.

\subsection{Rafting}

One of the earliest studies on rafting considering elastic stresses is due to Pineau~\cite{Pineau1977}.
The study of precipitate shape evolution and symmetry breaking transitions of Johnson and 
Cahn~\cite{JohnsonCahn1984} is another pioneering early study on the effect of elastic stresses on particle
morphologies. Following these, there 
have  also been several studies on the shape evolution and stability of precipitates under applied 
stresses considering single~\cite{BerkenpasEtAl1986,Johnson1987,WangLi2004} and 
multiparticle~\cite{NabarroEtAl1996,RatelEtAl2006} scenarios. 

One of the difficulties with these analytical studies is the evaluation of elastic field 
for arbitrary shapes of precipitates
and taking into consideration the elastic moduli differences. However, based on these analytical
studies, it was shown that the driving force for rafting is proportional to (i) the elastic moduli
mismatch; (ii) the misfit; and (iii) the applied stress (when the elastic moduli mismatch is small). 
In fact, the sign of rafting, namely, if the precipitates coarsen perpendicular (called N-type rafting) 
or parallel (called P-type rafting) 
to applied uniaxial stress depends on the signs of these three quantities. Let $\delta$ be the ratio of 
the shear modulus of the precipitate to that of the matrix. Then, P (N) type rafting occurs when 
$\bm \sigma^A \bm \varepsilon^0 (1-\delta) < 0 (> 0)$ where $\bm \sigma^A$ is the applied stress 
(with tensile being positive) and $\varepsilon^0$ is the eigenstrain 
(which is assumed positive if the lattice parameter of the precipitate is larger than the precipitate)
for small $\delta$. Thus, changing the sign of any of these keeping the other two constant will switch 
the type of rafting. In addition, the differences in anisotropy and the Poisson's ratio of the two
phases as well as large deviations of $\delta$ from unity has a strong say 
on rafting~\cite{SchmidtGross}; however, in those cases the above rule breaks down.

Note that most of the analytical studies of 
rafting assume Hookean elasticity and are based on thermodynamic considerations  -- either by considering 
the elastic energies associated with different shapes of coherent precipitates or by considering
the chemical potential contours surrounding a misfitting precipitate.

The most complete analysis of the rafting problem (assuming purely elastic stresses) is 
due to Schmidt and Gross~\cite{SchmidtGross}. Schmidt and Gross consider the instantaneous chemical potential at different points of a given precipitate and use it to predict rafting behaviour. 
Here we summarise the approach of Schmidt and Gross by highlighting the key steps and the results: the
algebra is fairly detailed and we refer the interested  reader to~\cite{SchmidtGross}.

\begin{itemize} 

\item Consider a two phase material with a coherent, misfitting precipitates in a matrix, assuming both the matrix and 
precipitate phases to be Hookean elastic. The change in potential of the system (strain energy), $\delta W^{el}$, 
due to the migration of the matrix-precipitate interface by an amount $\delta l$ in the direction $\bm n$ (normal to the
precipitate-matrix interface into matrix from the precipitate) is calculated using the energy-momentum tensor of Eshelby:
\begin{equation}
\delta W^{el} = - \int \tau_n \delta l dA 
\end{equation}
where $\tau_n = \bm n \cdot [\bm P] \bm n$ is the driving force with $[\bm P]$ is the jump in the Eshelby energy-momentum tensor.

\item Using the traction and displacement continuity equations, and using Eshelby's classic result~\cite{Eshelby1957}, namely
that the total strain inside an inclusion is related to eigenstrain linearly through Eshelby tensor ($\bm S$), one can show that
\begin{equation}
\tau_n = \frac{1}{2} \bm \epsilon^0 : \bm \Xi : \bm \epsilon^0 
\end{equation}
where $\bm \Xi$ is a fourth rank tensor which is related to elastic moduli of the two phases as follows (as hence has the same
symmetry as moduli tensor):
\begin{equation}
\bm \Xi = ([\bm C] \bm S + \bm C^p)^{T} \bm \gamma (\bm n) ([\bm C] \bm S + \bm C^p) - \bm \Lambda
\end{equation}
where
\begin{equation}
\bm \Lambda = \bm C^p + \bm C^p [\bm C] \bm C^p
\end{equation}
and
\begin{equation}
\bm \Gamma (\bm n) = [\bm C]^{-1} - \bm n \otimes \bm \Omega^{-1} (\bm n) \otimes \bm n
\end{equation}
with $\otimes$ is the tensor product (outer product) and $\bm \Omega$ is the acoustic tensor of the matrix: $\Omega_{ik} = C^m_{ijkl} n_j n_l$.
Note that $\bm \Xi$ is only the function of elastic moduli and the shape of the precipitate. 

\item Consider the two phase system to be under an externally applied stress. Schmidt and Gross show that 
this problem is equivalent to the problem above (that is, without applied stress) albeit with a 
modified (``equivalent") eigenstrain ($(\bm \epsilon^0)^{\star}$): 
\begin{equation}
(\bm \epsilon^0)^{\star} = \bm \epsilon^0 - [\bm C^{-1}] {\bm \sigma^{\infty}}
\end{equation}
where $\bm \sigma^{\infty}$ is the applied stress, can be defined;
This idea is equivalent eigenstrain is very similar to that of Eshelby~\cite{Eshelby1957};
however, while Eshelby reduces the problem to one of homogeneous inclusion using his equivalent eigenstrain, 
in Schmidt and Gross's case, the inclusion remains inhomogeneous.

\item Schmidt and Gross also manage to show that in case the precipitate volume remains constant and only shape changes, the
driving force for the modified problem is the same as the original problem. These results are valid for more than one precipitate;
it is independent of the geometry of the system; and, it is valid for arbitrary inclusion shapes.

\item Once the equivalent inclusion is known, the problem can be solved using the $\bm \Xi$ tensor for the modified eigenstrain. In case
the precipitate volume fraction does not change, the modified $\tau_n$ calculated using the equivalent eigenstrain (and hence the modified
$\bm \Xi$) gives the change in potential. 

\item At this stage, to proceed further, it becomes necessary to assume some symmetry for the precipitate as well as their distribution, 
and calculate the change in potential for different elongations -- as shown schematically.  For a spontaneous process, the change in 
potential should be negative. Using this condition, maps of normalised applied stress and normalised elastic moduli ratio indicating 
the different regions of P or N type rafting are obtained (as shown in schematic).

\end{itemize}

\begin{figure}[thpb]
\begin{center}
\resizebox{3in}{!}{\rotatebox{0}
{\includegraphics{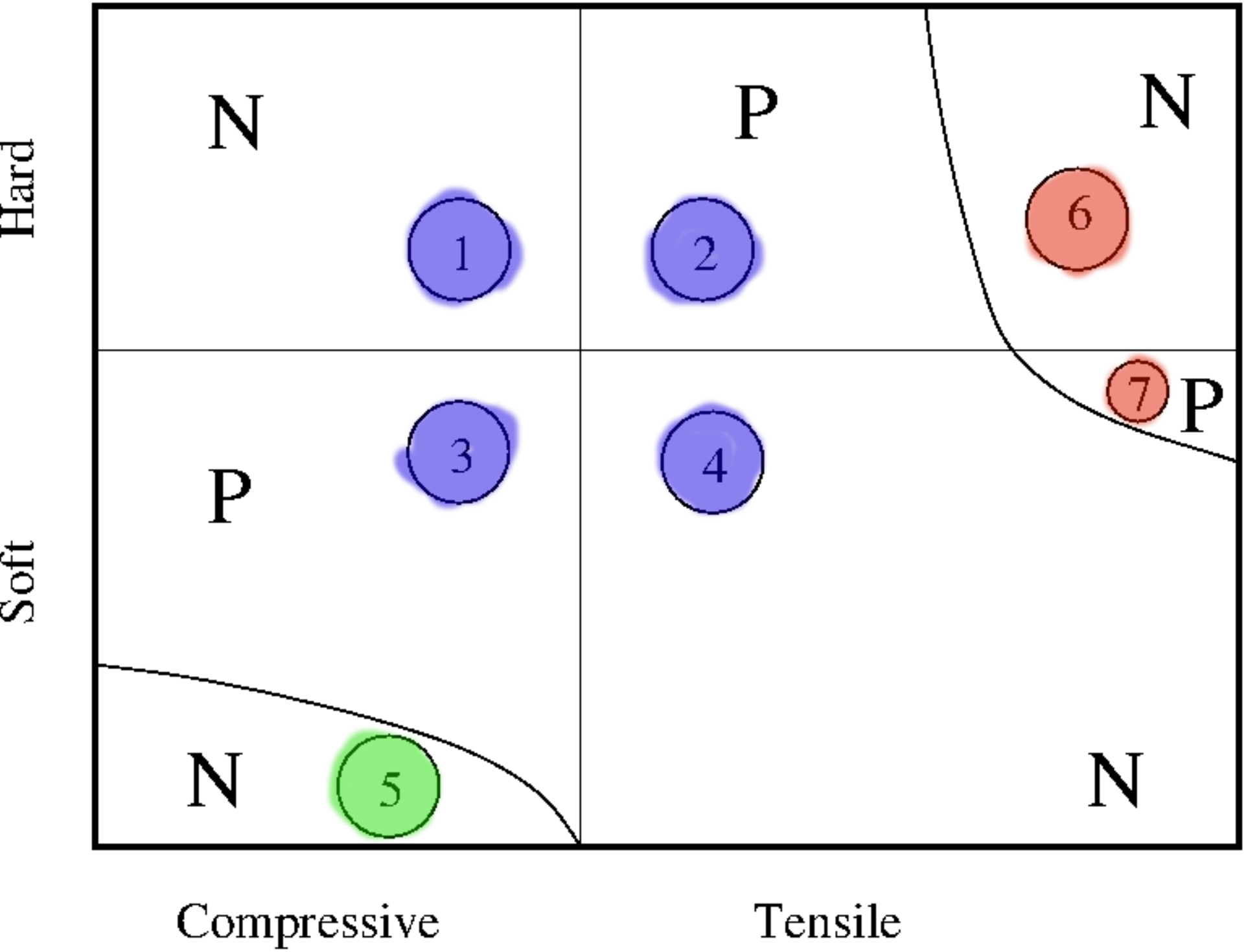}}}
\caption{Schematic rafting behaviour based on the analysis of Schmidt
and Gross, assuming a positive dilatational eigenstrain and an applied
stress along the $x$-axis on particles with four-fold symmetry; the red regions (6 and 7) require the
elastic anisotropy and Poisson's ratio of the two phases to be different; the green region (5) corresponds
to $\delta \ll 1$; the blue regions (1, 2, 3, and 4) are where the P and N rule is valid. The figure is 
based on~\cite{SchmidtGross} and adapted from~\cite{GuruThesis}.}
\label{SG_schematic}
\end{center}
\end{figure}

\subsection{Asaro-Tiller-Grinfeld instabilities}

There are two approaches to the study of ATG instabilities; one is a perturbation analysis and the other is variational approach. 
While the variational approach results 
are stronger in the sense that 
the morphology that minimizes the free energy is identified using it, the perturbation analysis 
will give the morphology 
also taking into account the kinetics. There are morphological instabilities such as dendritic morphologies formed during solidification 
which are a result of kinetics (how fast can the solidification front can move) and actually increase the interfacial free energy. 
Thus, both approaches are needed for a complete understanding of the morphological instabilities. In addition, since the ATG 
instabilities are of two types, namely, static and dynamic, while variational approaches are ideal for static studies, the perturbative 
approach can take the growth into account while analysing the stability.

There are several nice reviews which discuss the elastic stress effects, from the point of view of applications, 
on surface instabilities~\cite{MullerSaul2004}, on semiconductor heteroepitaxy~\cite{Teichert2002} and 
epitaxial growth~\cite{Emmerich2003}, during crystal growth by atomic and  molecular beams~\cite{PolitiEtAl2000},  and, 
on wrinkling of surfaces in soft materials~\cite{LiEtAl2012}. The instability analysis itself is summarised in several
articles for several different scenarios: see, for example~\cite{Grinfeld1987,Srolovitz1989,Grinfeld1991,Heidug1991,
Grinfeld1993,Grinfeld1993a,Grinfeld1993b,Grinfeld1993c,FreundJonsdottir1993,SpencerEtAl1993,HeidugLeroy1994,
LeroyHeidug1994, Gao1994,
Freund1995,SridharRickmanSrolovitz1997,SridharEtAl1997,Norris1999,Danescu2001,
SavinaVoorheesDavis2004,MisbahEtAl2004,Colin2004,ColinGrilhe2004,KornevSrolovitz2004,
YangSong2006,Yang2006,Colin2007,Mueller2008}. These papers can be classified in many different ways; for example, 
they can be classified based on the phases involved as solid-solid (for example, multilayers of solids), 
solid-liquid (for example, minerals in contact with stressed solids) and solid-vapour (for example, elastic half-spaces, 
plates and thin films); or, they can be classified based on the mechanism assumed 
(volume diffusion or surface diffusion or evaporation-condensation); or, they can be classified based 
on  the assumptions about the elastic properties (isotropic or orthotropic);
or, they can be classified based on the geometry they assume (cylindrical pore, spherical cavity, thin film and so on);
or, they can be classified based on the source of elastic strain (applied stresses, pressure in fluid,
coherency strains, imposed strains due to epitaxy and so on);
or, they can be classified based on whether they consider linear or nonlinear effects;
or, they can be classified based on the approach they take, namely the perturbative approach of Asaro-Tiller or the 
variational approach of Grinfeld. In this section, for the same of completeness, we briefly summarise the steps involved 
in the perturbative (using solid-solid, thin film geometry example~\cite{SridharRickmanSrolovitz1997}) and variational 
(using solid-liquid, curved solid in contact with fluid geometry example~\cite{Heidug1991}) approaches.

\subsubsection{Variational approach}

The variational approach pioneered by Gibbs, namely, extremising the relevant free energy functional, is used to study the stability
of stressed solid in contact with the second, compliant phase (be it solid, liquid or vapour), is described in this section (with specific
reference to stressed solid in contact with a fluid). As noted by Heidug~\cite{Heidug1991}, by 
demanding that, at equilibrium, there should be no production of entropy, one can derive the condition of chemical
equilibrium at the interface as a jump condition, as done by Lehner and Bataille~\cite{LehnerBataille198485}. This entropy production
approach shows that the conditions derived using Gibbsian approach are generic and are independent of the specific constitutive behaviour
assumed for the bulk phases or the loading configuration that the solid is subjected to. 

\begin{figure}[thpb]
\begin{center}
\resizebox{4in}{!}{\rotatebox{0}
{\includegraphics{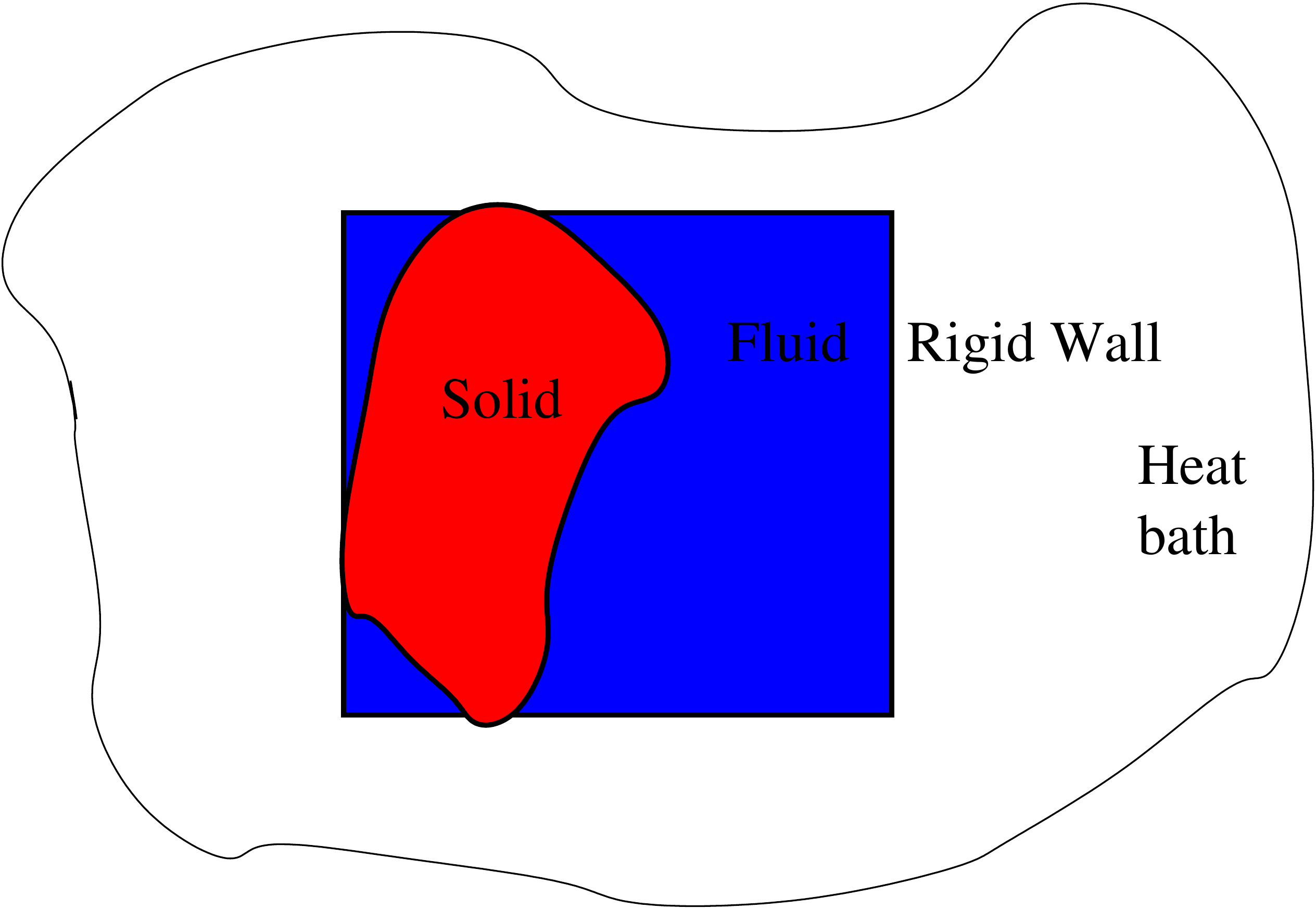}}}
\caption{Schematic of solid-liquid system in reference configuration; based on~\cite{Heidug1991}.} 
\label{SchematicHeidug}
\end{center}
\end{figure}

Let us consider the equilibrium of a solid-fluid system as shown in the schematic Fig.~\ref{SchematicHeidug}; 
the system is at constant
temperature and is enclosed by rigid boundaries. The equilibrium of such a system is determined by the minimization of the
(Helmholtz) free energy
\begin{equation}
\Psi = \int_{R_s+R_f} \psi dv + \int_{\Sigma} \hat{\psi} da
\end{equation} 
where the first term is the bulk free energy integrated over the solid ($R_s$) and liquid ($R_f$) volumes (in the reference state at equilibrium) 
and the second term is integrated over the interface area; the minimization has to be carried out subject to the conservation of the solvent 
(denoted by D) and solute (denoted by S) mass:
\begin{equation}
\int_{R_s+R_f} \rho^S dv + \int_{\Sigma} \hat{\rho}^S da = \mathrm{Constant}
\end{equation}
\begin{equation}
\int_{R_s+R_f} \rho^D dv + \int_{\Sigma} \hat{\rho}^D da = \mathrm{Constant}
\end{equation}

Thus, the problem we are considering is one of constrained minimization. So, we introduce the (undetermined) Lagrange multipliers $\lambda^S$ and $\lambda^D$ corresponding to the two constraints, and write the new functional to be optimized as follows:
\begin{equation}
\Phi = \Psi +  \lambda^S \left[\int_{R_s+R_f} \rho^S dv + \int_{\Sigma} \hat{\rho}^S da \right] + \lambda^D \left[ \int_{R_s+R_f} \rho^D dv + \int_{\Sigma} \hat{\rho}^D da \right]
\end{equation}
The minimization is achieved when the first variation $\delta \Phi$ is zero and the second variation is positive. The first variation
leads to local equilibrium conditions -- which, in this case, are as follows:
\begin{itemize}
\item The chemical potential for the solvent and the solute are uniform;

\item In solid and the fluid, the relevant equations of mechanical equilibrium is satisfied; that is, solid supports non-hydrostatic stress and in fluids the stress state is hydrostatic;

\item At the interface, force balance for stressed membranes is satisfied; namely, capillary equilibrium for solid-fluid interface is satisfied; 
and at the interface, the shear stresses in the solid are balanced by surface tension.

\end{itemize}

The key piece in this derivation is the identification appropriate allowed variations. 
Specifically, (i) the allowed variations should be such that there are no displacements at the system boundary 
(since we assumed it to be rigid); (ii) the allowed variations should be such that the displacements
at the interface have no discontinuity; (iii) in the solid phase, the allowed variations are such that their 
gradient is the same
as the variation of the deformation gradient; (iv) the allowed variations are such that the displacements at 
the interface, when decomposed 
into the normal and tangential components, give rise to tangential components that are compatible with the 
(Gaussian) surface parameters; 
and (v) the allowed variations are such that the interface velocity and the rate of change of interface 
metric that it gives rise to are
compatible.   

The second variation gives the stability criterion; it can be shown that stability
demands that (i) the stress on the solid at the interface should be hydrostatic and equal to fluid pressure; and (ii) 
either the Gibbs surface energy vanishes or that the interface is flat~\cite{Heidug1991}. 
Thus, phase equilibrium at non-hydrostatically stressed, curved solid-fluid interfaces is not stable.

\subsubsection{Perturbative approach}

\begin{figure}[thpb]
\begin{center}
\resizebox{4in}{!}{\rotatebox{0}
{\includegraphics{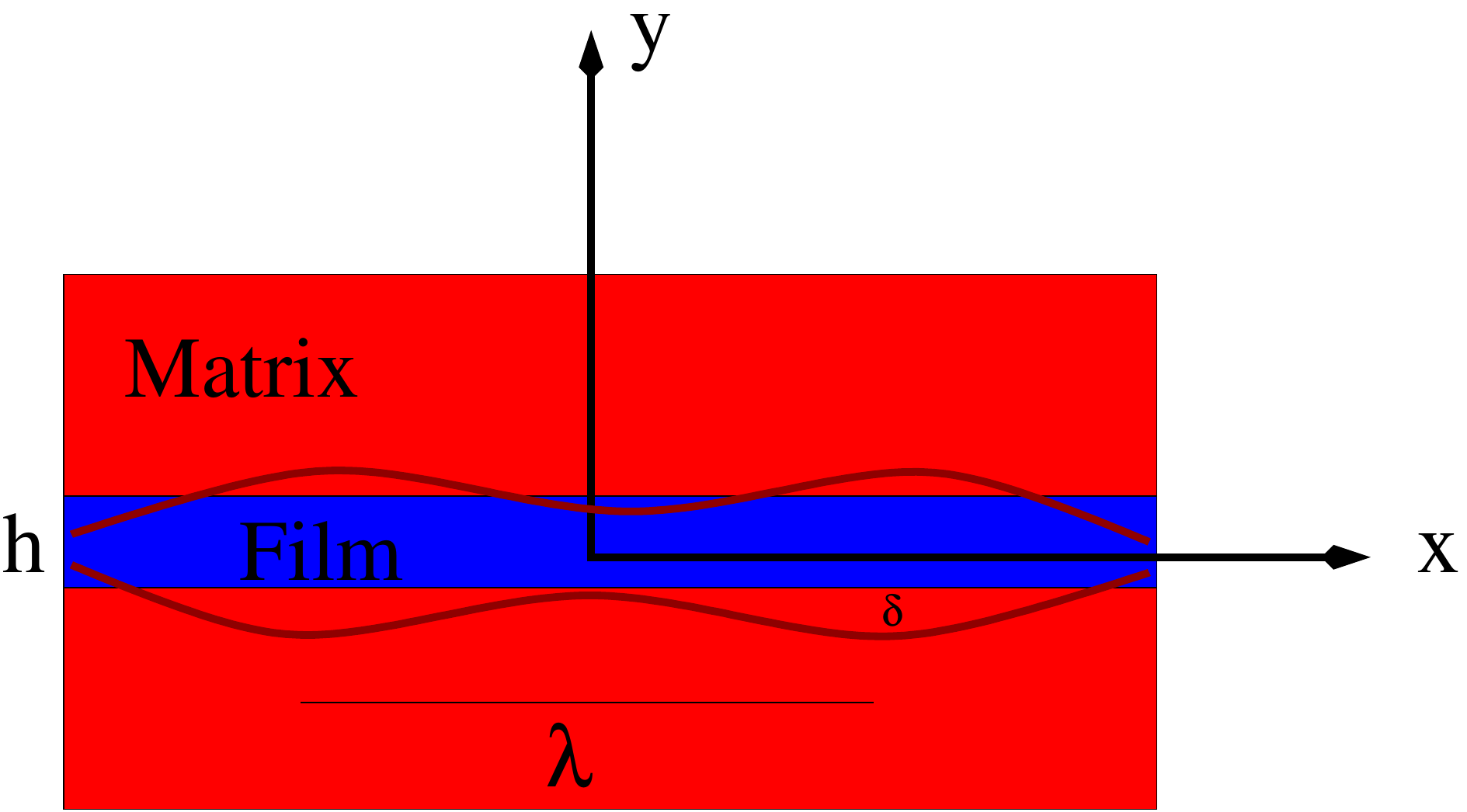}}}
\caption{Schematic of perturbation of the film-matrix interface; based on~\cite{SridharRickmanSrolovitz1997}.} 
\label{PerturbationGeometry}
\end{center}
\end{figure}

Let us consider a perturbation of the solid-solid interfaces as shown in Fig.~\ref{PerturbationGeometry} in an elastically stressed solid. 
The (sinusoidal) perturbed interface profile is described by $y_i = \pm [\frac{h}{2} + \delta \cos{(kx)}]$ where $\delta$ is the amplitude of
perturbation of wavelength $\lambda (= \frac{2 \pi}{k})$ where $k$ is the wavenumber and $h$ is the height of the film as shown in the figure.
We assume $\delta k \ll 1$; that is, the interface profile is such that its slope is very small everywhere; this assumption is what makes
the analysis perturbative.

Let $\mu_0$ be the chemical potential of the interface when it is flat and let $\mu$ be the chemical potential along the interface
when the interface is perturbed. Let $\gamma$ be the (isotropic) interfacial energy. Then, 
\begin{equation}
\mu - \mu_0 = \Omega \left( \kappa \gamma + [W]_-^+ - \mathbf{T} \cdot \left[ \frac{\partial \mathbf{u}}{\partial n} \right]_-^+\right)
\end{equation}
where $\kappa$ is the interface curvature; $\Omega$ is the atomic volume; $[W]_-^+$ is the jump in strain energy density across the 
interface; $\mathbf{T}$ is the traction on the interface; $\partial \mathbf{u} / \partial n$ is the derivative of the total displacement 
field in the direction normal to the interface. The first term on the RHS is due to interfacial energy; and, the last two terms on the RHS 
which are due to elastic stresses and was derived by Eshelby using his energy-momentum tensor~\cite{Eshelby1975}.

Thus, for the perturbed geometry, the equation of mechanical equilibrium is to be solved (under appropriate boundary conditions) and 
using the elastic solution obtained the second and third terms are to be evaluated. It is not possible to do this analytically. 
However, since we have assumed small slope for the interface everywhere, it is sufficient to obtain these quantities to first order 
in $\delta k$ and such an approximate expression can be obtained - see these notes for the MAPLE\texttrademark ~script
which can be used to obtain the elastic solutions~\cite{GuruATGNotes}. In addition, the interfacial curvature can also be shown, to first 
order in $\delta k$, to be $\delta k^2 \cos(kx)$.

Let F be the force on atoms at the interface. It is relationship to the chemical potential is as follows: $F = - \frac{\partial \mu}{\partial s}$
where $s$ is the distance along the interface (interfacial arc). By Fick's first law, the atomic flux is proportional to the force
and the proportionality constant is the mobility $M$; that is, $J = M F$. The mobility is related to the diffusivity and interface width
$\eta$ though $M = D \eta /(\Omega k_B T)$ where $k_B$ is the Boltzmann constant and T is the absolute 
temperature. Once the flux is given, 
using the conservation of mass, the velocity of the interface ($\nu$) can be calculated as
\begin{equation}
\nu = -\Omega \frac{\partial J}{\partial s} = M \Omega \frac{\partial^2 \mu}{\partial s^2}
\end{equation}
However, the velocity can also be calculated to first approximation, by differentiating the interface 
profile with time, and hence
\begin{equation}
\nu \approx \frac{\partial \delta}{\partial t} \cos(kx) = M \Omega \frac{\partial^2 \mu}{\partial s^2}
\end{equation}

The solution of this equation (since the RHS term can be shown to be sinusoidal) is
\begin{equation}
\delta (\tau)  = \delta(0) \exp{(\phi \tau)} 
\end{equation}
where  both the time, $\tau$ and growth rate $\phi$, are non-dimensional.

By solving the elastic problem under different boundary conditions and assuming different diffusion mechanisms,
Sridhar et al~\cite{SridharRickmanSrolovitz1997} have shown that there are two possibilities of break-up
for films -- namely, symmetric and anti-symmetric. In fact, Sridhar et al~\cite{SridharRickmanSrolovitz1997}
give stability diagrams indicating the parameter ranges of break-up and the type of break-up.

\section{Phase-field models} \label{PhaseFieldModels}

The models used for the study of elastic stress induced instabilities can be very broadly classified as atomistic models and 
continuum models. The Discrete Atom Method (DAM) is an example of the atomistic method. On the other hand, the continuum 
models can further be broadly classified as sharp interface and diffuse interface models. Most of the theoretical studies 
described in the previous section, for example, are sharp interface models. There are numerical implementations (at times 
based on the finite element method) of these sharp interface models (see for example, ~\cite{RamasubramaniamShenoy}). 
However, for the study of microstructural instabilities, as we show below, the diffuse interface models are the most ideal.

In the diffuse interface models, the microstructure is described using field variables (that is, variables which are defined 
for all space points at all times) and their derivatives. These field variables are also called order parameters. The order 
parameters typically take a constant value in the bulk phases and change from one bulk value to another in the interface 
region. Thus, the interface is defined as the region over which the order parameter changes. Hence, these models are called 
diffuse interface models. In these models, the bulk phases are defined by the constant value that the field variable takes 
inside of them. Hence, these models are also called phase field models -- to indicate that different phases are denoted by 
the field variables taking specific values.

There are several different ways in which one can understand phase field models. Here we list a few of these viewpoints -- though, 
these viewpoints are not exclusive. 

Phase field models can be thought of as a mathematical strategy to find solutions for hard-to-solve sharp interface models~\cite{CaginalpSocolovsky1989,CaginalpSocolovsky1991,Borst2008}. 
In such a scenario, we artificially assume a width to an interface, which, in reality, a plane of zero width. 
Such artificial, diffuse interface allows us to solve the resultant partial differential equations fairly easily. 
In such a viewpoint, the attempt is always to show that in the limit of the interface width going to zero, we obtain the 
corresponding sharp interface models and hence, in the limit of the interface width going to zero, we obtain the solution 
to the sharp interface problem from the corresponding diffuse interface solution. This viewpoint can be considered as a purely 
mathematical viewpoint because many physical interfaces are indeed diffuse. 
 
Phase field models can also be thought of as partial differential equations which lead to interesting patterns as solutions. In this
viewpoint, which is also relatively mathematical, the emphasis is on the solutions obtained. A classic example of this viewpoint is
the attempt of Alan Turing~\cite{Turing} to look at pattern formation (what he called as chemical morphogenesis) as a reaction-diffusion equation.

Another prominent viewpoint is to think of phase field models as continuum models (derived from statistical physics)
that lead to interesting patterns as solutions~\cite{HohenbergHalperin1977,ChaikinLubensky}. 
As in the case of biological pattern formation models, in this viewpoint also, the emphasis 
is on the solutions obtained. However, here an attempt is also made to connect the patterns 
obtained to the underlying physical 
processes and statistical mechanics. To that extent, this viewpoint is more physics based. 

The viewpoint (which we will take in this review, and which we call as the materials science based approach) is to consider 
phase field models as non-classical diffusion equations. In this viewpoint, we begin by modifying the classical thermodynamics
of materials. In classical thermodynamics, the interface width is arbitrarily assumed (typically, zero -- though not always);
and, in calculations (such as in phase diagram construction) the interface contribution is further assumed to be 
negligible.
If we incorporate this interface contribution and allow the system to choose the interface width consistent with the 
imposed
thermodynamic variables and constraints, the resultant non-classical thermodynamics (along with certain constitutive 
laws), as we show below,
leads to equations which are non-linear diffusion equations. This was the approach pioneered by Cahn and Hilliard in formulating the 
Cahn-Hilliard equation~\cite{Cahn-Hilliard} and explained very lucidly in his pedagogical article by 
Hilliard~\cite{HilliardPhaseTransArticle}.

The contribution of Cahn and Hilliard (based on the earlier atomistic studies of Hillert) is to
show that this wavelength limit is set by interfacial energy of the incipient interfaces in the system and to get this limit
out of the model, the interfacial energy contribution should be incorporated into the free energy. In the next subsection, we will indicate
the modification to the free energy and the derivation of the Cahn-Hilliard equation. This is one of the two canonical phase 
field equations. In the following subsection, we will indicate the other canonical phase field equation called 
Allen-Cahn equation (or, sometimes Time Dependent Ginzburg-Landau (TDGL) equation, or, simply, Ginzburg-Landau equation). 
All phase field models can be thought
of as a combination of these two models. These two subsections will also set the stage for us to describe the process 
of formulating phase field models in a more abstract fashion.

\subsection{Cahn-Hilliard equation}

The classical Gibbs free energy in a binary alloy is a function of composition.  The derivative of this free energy with respect to the
A or B atoms gives the corresponding chemical potential. Using the chemical potential in the modified Fick's first law in combination with 
law of conservation of mass leads to the classical diffusion equation. 

Cahn and Hilliard showed that in order to account for interfacial energy contribution to the free energy, the free energy should be made
a function of not just composition but also its derivatives (spatial, such as gradient, curvature, aberration and so on). This implies that
the free energy is not a function but functional. Specifically, in the case of a binary alloy (assuming isotropy or cubic anisotropy), Cahn and Hilliard showed that the free energy functional is of the form
\begin{equation}
G(c,\nabla c,...) = N_V \int_{V} dV [f_0(c) + K |\nabla c|^2]
\end{equation}
where $K$ is the gradient energy coefficient (assumed constant), and $f_0(c)$ is the bulk free energy density.

Since the free energy is a functional, the chemical potential is given by the variational derivative of the free energy functional (the Euler-Lagrange equation):
\begin{equation}
\mu = \frac{\delta (G/N_V)}{\delta c} = \frac{\partial f_0}{\partial c} - K \nabla^2 c
\end{equation}
Using this chemical potential, we can define the flux as
\begin{equation}
\mathbf{J} = - M \nabla \left[ \frac{\partial f_0}{\partial c} - K \nabla^2 c \right]
\end{equation}
This flux, along with the conservation of mass, leads to
\begin{equation}
\frac{\partial c}{\partial t} = \nabla M \nabla \left[ \frac{\partial f_0}{\partial c} - K \nabla^2 c \right]
\end{equation}

Thus, the Cahn-Hilliard equation is given as
\begin{equation}
\frac{\partial c}{\partial t} = M \frac{\partial f_0}{\partial c} \nabla^2 c - M K \nabla^4 c 
\end{equation}
where we have assumed the mobility to be a constant.

Comparing this equation with the classical diffusion equation, we see that there is an extra non-linear term ($\nabla^4 c$).

\subsection{Allen-Cahn equation}

The Allen-Cahn equation can be derived in a very similar fashion. Let us assume that the microstructure 
is described using
an order parameter $\phi$. For simplicity, we assume that the $\phi$ parameter takes to distinct values 
(say, zero and unity)
in two phases and takes values between zero and unity in the interface region. Let $f_0(\phi)$ be a 
double-well potential with
minima at zero and unity. Let us consider a free energy functional that describes the thermodynamics of the system:
\begin{equation}
G(\phi,\nabla \phi,...) = N_V \int_{V} dV [f_0(\phi) + K |\nabla \phi|^2]
\end{equation}
where $K$ is the gradient energy coefficient (assumed constant).

In this case also, we can define a chemical potential:
\begin{equation}
\mu = \frac{\delta (G/N_V)}{\delta \phi} = \frac{\partial f_0}{\partial \phi} - K \nabla^2 \phi
\end{equation}

In the Allen-Cahn case, we assume that the order parameter is not a conserved quantity 
(unlike the case of composition where $\frac{d}{dt} \left[ \int c dV \right] = 0$). Hence, 
we assume the following constitutive law for the rate of change of order parameter:
\begin{equation}
\frac{\partial \phi}{\partial t} = - L \mu
\end{equation}
where $L$ is the relaxation parameter~\cite{JOMArticleCarterEtAl}. 

Hence, one obtains the Allen-Cahn equation as
\begin{equation}
\frac{\partial \phi}{\partial t} =  L K \nabla^2 \phi - L \frac{\partial f_0}{\partial \phi}
\end{equation}
This equation is also known as TDGL equation or reaction-diffusion equation since it is very similar to diffusion 
equation except for the non-linear polynomial in $\phi$ which is like a source/sink term due to 
chemical reactions. 

\subsection{Incorporating elastic stress effects}

Since we concentrate on the elastic stress induced microstructural instabilities in this review, 
the incorporation of elastic stress effects into the formulation is a key step. The incorporation 
of elastic stress effects into the phase field models is achieved by adding the elastic
energy $F^{el} = \frac{1}{2} \int \bm \sigma^{el} \bm \varepsilon^{el}$ to the free energy functional.
The elastic stress and strain fields are obtained by solving the equation of mechanical equilibrium.

In all these models, since the time scales of elastic relaxation are much larger than the diffusional
time scales, the phase field equations and the equation of mechanical equilibrium are solved 
sequentially, assuming that for any given order parameter field, the elastic fields equilibrate 
instantaneously; in addition,  
the eigenstrain or the elastic moduli or both are slaved to the order parameter. 
The resultant equation of mechanical 
equilibrium is solved with either imposed strain or applied traction boundary conditions.

In models of microstructure evolution, it is very common to assume that the domain of computation 
is a representative volume element; in other words, it is common to use periodic boundary conditions. 

As discussed earlier, in most cases of interest, the elastic moduli are anisotropic (at least cubic) 
and inhomogeneous; there are eigenstrains (primarily,
due to coherency) and applied stresses. Hence, slaving the eigenstrains and elastic moduli to the 
order parameters makes solving the equation of mechanical equilibrium becomes one of homogenisation problem. 

The equation of mechanical equilibrium in coherent, anisotropic and inhomogeneous systems can, 
of course, be solved using finite element techniques; there are several studies which do use 
finite element techniques. However, in phase field models, in certain cases, like,
for example, in the case of spinodal decomposition, such finite element techniques can 
become very difficult from an implementation /  computational cost point of view since to 
capture interface very fine meshing is needed and the microstructure is full of interfaces. 
In addition, as the microstructure evolves, the mesh also needs frequent updating. Hence, 
Fourier transform based spectral techniques are very widely used and are quite successful.

\subsection{Formulation}

As noted at the beginning of this article, microstructure is nothing but the size, 
shape and distribution of interfaces; specifically,
when we are studying elastic stress driven microstructural instabilities, 
we are interested in the formation, disappearance, break-up and/or 
merger of interfaces in elastically stressed systems. Thus, any model 
that we formulate to study stress driven microstructural 
instabilities should be capable of describing the microstructure (geometry or topology), 
its energetics (thermodynamics), and kinetics; in
addition, since we are interested in stress driven microstructural changes, 
our energetics should include the strain energy, which, in turn,
should be calculated using the appropriate physics.

The two canonical phase field models that we discussed above, namely the 
Cahn-Hilliard equation for systems with conserved order parameters, and
Allen-Cahn equations for systems with non-conserved order parameters 
leads to non-linear diffusion equations (from a mathematical point of view; 
note that physically, while Cahn-Hilliard is actually a modified diffusion 
equation, Allen-Cahn is not). However, from the derivation of these
two equations it is clear that the general formulation of phase field models 
(which are nothing but a combination of these two types of equations)
consists of the following steps:

\begin{itemize}

\item {\bf Description of microstructure (the geometry/topology)}

The first step in formulating a phase field model is to identify the 
order parameter that describes the microstructure. The order parameter can
be a conserved quantity (such as composition) or a non-conserved quantity 
(such as ordered domain of a given type). 

\item {\bf Thermodynamics}

The second step is to describe the thermodynamics of the system. We do this by 
defining the free energy or entropy functional; these
thermodynamic functionals are given in terms of the order parameters and their 
spatial derivatives. In our viewpoint, it is such
thermodynamic description (in terms of functionals) that make phase field 
models what they are. If the thermodynamics is described using 
classical free energy functions (without (at the least) the gradient terms), 
the resultant partial differential equations will lead to
sharp interface and not diffuse-interface description.

\item {\bf Kinetics}

Given a free energy functional, one can define the chemical potential. 
In terms of the chemical potential, then, there are two constitutive laws
that we use which introduce the kinetics -- how fast or slow the system 
relaxes to its equilibrium (since, for equilibrium, the Euler-Lagrange
equations should equal zero): in the case of Cahn-Hilliard equation, it 
is the mobility and in the case of Allen-Cahn equation, it is the
relaxation parameter.

\item {\bf Conservation laws and other physics}

As in the case of Cahn-Hilliard equation, after the introduction of 
kinetics, we may have to impose any other relevant conservation laws such
as conservation of mass, energy and charge. In addition, in the case 
of elastically stressed systems that we discuss in this paper,
the free energy will also consist of the elastic energy terms. 
These elastic energy terms are to be computed using the relevant 
physics: that
is, the equation of mechanical equilibrium should be solved under 
appropriate boundary conditions and the resultant stress and strain fields 
along with applied stresses (if any) should be used to compute the 
elastic energy term. Similar process has to be carried out if the free
energy contains electric, magnetic or any such other energy terms that 
are relevant~\cite{CarterElectricalPaper}.

\end{itemize}

At this point, it is to be emphasised that phase field modelling is a methodology; 
for example, for the same problem, there could be
more than one description in terms of order parameters and the energetics; 
this depends on the level of detail that we wish to incorporate. 
There is no ``the" phase field model for any given problem. A good examples 
of this, in our context is to think of phase field models
for elastically stressed systems: one can consider scalar order parameters 
and make the eigenstrains slaves of such order parameters (which
is the more common approach); however, one can also think of the strains 
as the order parameters and evolve them by writing corresponding 
free energies (if we can). Similarly, in the case of Ni-base superalloys,
for example, if anti-phase boundary (APB) related physics is not important,
they can be modelled using a single composition order parameter. However, if
APBs are important, in addition, one should introduce three additional non-conserved
order parameters so that the four variants can be completely described.

From the description above, yet another viewpoint on phase field models emerges. 
In this viewpoint, phase field models are partial differential 
equations that describe the evolution of order parameters that describe microstructures; 
the order parameters are field variables; they take 
constant values in the bulk and change in the interface region and thus, highlight the 
interfaces and hence help us understand the formation
of microstructures and their evolution.

There are two characteristics of the solutions of the phase field equations which 
are very important. The first is of course the diffuse interface
solution (which is a direct consequence of the inclusion of gradient terms); 
this means that there are no discontinuities in the domain and hence
there is no need for tracking of interfaces (to impose jump conditions for example). 
This makes the numerical solutions much easier and also
makes the processes of dealing with formation, disappearance, merger and splitting of 
interfaces fairly easy. The second is that the interface 
physics (that which is relatable to the interfacial energy primarily and not so 
much interface structure -- such as Gibbs-Thomson effect, for 
example) are automatically taken into account in these models. There is no 
need to incorporate them separately as is sometimes done in sharp 
interface models.

\subsection{Parameters, non-dimensionalisation and numerical implementations}

The parameters that enter the phase field model for stress induced microstructural 
evolution are the following: 

\begin{itemize}

\item {\bf Related to the thermodynamics of the system} 

Bulk free energy density and the gradient energy coefficients;

\item {\bf Related to the kinetics}

The mobilities and relaxation parameters; and,

\item {\bf Related to equation of mechanical equilibrium} 

The eigenstrains and the elastic moduli along with their dependence on the order 
parameters; and applied stresses or imposed strains. 

\end{itemize}

In addition, there are numerical implementation related parameters that enter the 
calculations such as the domain size, the spatial grid
size, and the time-steps of integration. Finally, in the numerical solution of 
the equation of mechanical equilibrium, there are magic numbers
that enter the calculation such as convergence criterion for elastic fields.

In general, while solving equations on a computer, it is preferable that the 
equations are non-dimensionalised. This makes the computations robust
and can help avoid repetitions in calculations. A careful choice of non-dimensionalisation 
is also essential to carry out meaningful simulations. Let
us consider a typical microstructure in which the interface is a few lattice 
parameters wide, say 1 nm or so. To capture the interface in the
numerical model, we need nearly six to eight mesh points. Assuming that 
the interface will be captured using eight mesh points, one can see
that the spatial discretization corresponds to about 1.25 \AA. This can, 
at times, be very restrictive. Appropriate non-dimensionalisation
can help overcome this problem as explained below.

In the classical Cahn-Hilliard model, there are two interfacial parameters, 
namely, the width and energy. One can use the parameters associated with
the thermodynamics, namely, the energy barrier between the two phases in the 
bulk free energy density and the gradient energy coefficient by non-dimensionalising 
these quantities using the interfacial energy and interfacial width, while the kinetic 
parameters are used to non-dimensionalise
time. Such non-dimensionalisation helps us study bigger systems and is described in 
detail in~\cite{GuruThesis}. 

At this point, it is also clear that there are several quantities which enter the phase 
field models, which are difficult to measure experimentally.
For example, the coherency strains, the moduli and their dependence on composition, 
the interfacial energy, and the mobility are difficult to 
measure experimentally, though, reliable measurements of the bulk free energies 
(in the form of CALPHAD data, for example) are available;
in some cases, diffusivity is also available. 

In Cahn and Hilliard's work, in addition to connecting their bulk free energy density 
term to a regular solution model, they have also attempted to
relate the gradient energy coefficient to the bond energies. However, 
these attempts are not very successful. Hence, in most phase field models, 
at present, it is far more easier (and reliable) to get trends than to get actual 
quantitative information though attempts are being made to make the phase field
models more quantitative; see for one of the early attempts~\cite{ZhuEtAl2004}.
 
Finally, the phase field equations can be solved using any of the available numerical 
techniques: finite difference, finite volume, finite element (see for 
example~\cite{BarrettEtAl2005,ZaeemMesarovic2010} and references
therein), boundary integral method~\cite{JouEtAl1997} and spectral techniques~\cite{ChenShen1998}. 

As noted by~\cite{ZaeemMesarovic2010}, though using finite element method
for Allen-Cahn equations is easy, for Cahn-Hilliard method higher order 
interpolation methods are needed. Simulating local mass flux and handling topological 
singularities are very difficult in boundary integral methods; the method might also require
preconditioners to solve the resulting system of equations~\cite{JouEtAl1997}.
 
Khachaturyan, Chen and their co-workers have pioneered the use of spectral techniques.
Spectral techniques have several advantages; they automatically incorporate periodic 
boundary conditions which are the relevant boundary conditions for the representative
volume elements. Even though they do not convert the partial differential equation
into an ordinary differential equation since phase field models consist of non-linear terms,
they still can be implemented using semi-implicit techniques and hence allow for
relatively larger time steps and they also result in spectral accuracy. 
They can also handle higher order derivatives very well.
One disadvantage of spectral techniques, of course, is that boundary conditions
other than periodic boundary condition is difficult to incorporate which can be done
fairly easily in finite difference techniques for example.

In terms of numerical implementation, even though finite difference techniques are more involved,
they can be easily parallelised unlike spectral techniques. However, in recent times, the GPU based 
parallelisation of FFT (such as CUDAFFT) has given some advantage in terms of parallelization to spectral
techniques.
 
\subsection{Benchmarking against analytical solutions}

Several authors have carried out more formal asymptotic analysis for phase field models 
incorporating elastic stress effects; see Fried and Gurtin~\cite{FriedGurtin1994}, 
Leo et al~\cite{LeoEtAl1998} and Garcke and Kwak~\cite{GarckeKwak} for some representative
examples.

When it comes to numerical implementation of phase field models, benchmarking the numerical 
solutions obtained from phase field models against classic elastic solutions (in those 
cases where they are available) is very important.
Such benchmarking serves to show the numerical implementation is correct. In addition, 
though not as rigorous as the analysis of, for example, Garcke and Kwak, such benchmarking
can be thought of as an engineering approach to checking on the correctness of phase
field formulation. In this subsection, we list examples of such benchmarking from some of
our work; similar benchmarks have been reported by other authors too.

\begin{itemize}

\item Chirranjeevi et al~\cite{ChirranjeeviEtAl} have confirmed that the phase field
models does indeed show the symmetric and anti-symmetric break-up of films as predicated
by Sridhar et al~\cite{SridharRickmanSrolovitz1997}. Further, at the very early stages
of the break-up, the maximally growing wavelength compares well with the analytical solution~\cite{GuruThesis}.

\item Gururajan and Abinandanan~\cite{GuruAbiRafting2007} have obtained all the five regions
identified by Schmidt and Gross~\cite{SchmidtGross}. In addition, they have also verified that
the results from the phase field model compare well with Eshelby solution
for inclusions and inhomogeneities (including voids)~\cite{Mura}, and the homogeneous strain~\cite{Khachaturyan} for homogeneous alloys~\cite{GuruThesis}.

\item Mukherjee et al~\cite{RajdipEtAl} show that phase field models predict the curvature and coherency
driven Gibbs-Thomson effect very well; further, in 1-D, in systems with no coherency strains, the
growth rates are shown to agree well with the classical solutions of Frank~\cite{Frank} and Zener~\cite{Zener}.

\end{itemize} 

\subsection{Spinodal phase separation: suppression and promotion}

The phase field implementation of phase separation in elastically stressed systems have been many: see for example~\cite{OnukiNishimori1991,WangChenKhach1993,SaguiEtAl1994,ThomsonVoorhees1997,WangBanerjeeSuKhachaturyan,
LiChen1999,HuChen2001,ZhuChenShen2001,SeolEtAl2003,BoussinotEtAl2009}.
As we noted above, the presence of elastic stresses or strains tends to suppress spinodal. However, 
when the system does undergo spinodal, the composition modulations in elastically softer directions 
grow leading the phase separation that is anisotropic. 

If the system is elastically inhomogeneous, 
the harder phase becomes more compact but deviates from
spherical shape and takes shapes that are consistent with their elastic anisotropy; for example, 
in cubic systems they become cuboids; they also preferentially 
align along the elastically soft directions. In addition, the coarsening rates in such systems after phase 
separation is slow. Finally, as we show below, the compact precipitate phases might split; and, in 
the presence of applied stresses, they coarsen preferentially along certain directions.

However, if there are imposed strains on the system (for example, as in the case of an epitaxially grown 
thin film undergoing spinodal decomposition), then the elastic stresses can promote spinodal decomposition 
even outside  the chemical spinodal~\cite{LahiriEtAl}. In this section, using the
analytical solution derived in the previous section, we extend the analysis of Cahn 
and show that the coherent spinodal region extends beyond chemical spinodal.

\subsubsection{\label{computing_the_spinodal}Computing the spinodal}

Here we assume a regular solution model for computing the spinodal lines. 
According to Cahn\cite{Cahn1961}, the chemical spinodal is given by,
\begin{eqnarray}
\frac{\partial^{2} G}{\partial c^{2}}=0,
\label{spinodal:one}
\end{eqnarray} 
where,
\begin{eqnarray}
G=G_{A}(1-c)+G_B c+RT[c \ln{c}+(1-c) \ln{(1-c)}]+\nonumber \\
\Omega c(1-c), \nonumber \\
\label{spinodal:two}
\end{eqnarray}   
where $G$ is the molar Gibbs free energy, $R$ is the Universal Gas constant and $\Omega$ is the molar heat of mixing. The critical temperature is given by:
\begin{eqnarray}
T_{c}=\frac{\Omega}{2R}.
\label{spinodal:three}
\end{eqnarray}

For the coherent spinodal in the plane stress setting assuming isotropic elasticity, we have:
\begin{eqnarray}
\frac{\partial^{2} G}{\partial c^{2}}+Y\eta^{2}V=0,
\label{spinodal:four}
\end{eqnarray}
where the Young's modulus $Y$ is a function of homogeneous alloy composition and $V$ is the molar volume of the homogeneous alloy under question. 

Now we consider the situation where the elastic modulus of the system is composition dependent and the system is subjected to an applied homogeneous strain (constrained spinodal). We start out with a homogeneous alloy having a composition $c_{0}$, whose strain energy density is given by:
\begin{eqnarray}
W_{E}^{0}=\frac{1}{2}[Y_{A}+(Y_{B}-Y_{A})c_{0}]{e}^{2},
\label{eq_ana_ex:one}
\end{eqnarray}  
where we assume the lattice of an undecomposed 50:50 alloy to be the reference. In the presence of a composition modulation, the strain energy density becomes:
\begin{eqnarray}
W_{E}^{1}=\frac{1}{2}[Y_{A}+(Y_{B}-Y_{A})c]{[e-\eta (c-c_{0})]}^{2}\nonumber 
\label{eq_ana_ex:two}
\end{eqnarray}   
Now, assuming $c-c_0=Acos\beta x$ we get the expression of the total strain energy as:
\begin{eqnarray}
W=\int_\Omega [W_{E}^{1}-W_{E}^{0}] d\Omega=\nonumber \\
\frac{1}{2}\int_{\Omega}[Y_{A}\eta^{2}+3\eta^{2}(Y_{B}-Y_{A})c_{0}-2\eta e (Y_{B}-Y_{A})]{(c-c_{0})}^{2} d\Omega.\nonumber \\
\label{eq_ana_ex:four}
\end{eqnarray}  
This defines the \emph{constrained spinodal} as:
\begin{eqnarray}
\frac{\partial^{2}G}{\partial c^{2}}+[Y_{A}\eta^{2}+3\eta^{2}(Y_{B}-Y_{A})c_{0}-2\eta e (Y_{B}-Y_{A})]V=0\nonumber \\
\label{eq_ana_ex:five}
\end{eqnarray}
where $V$ denotes the molar volume.

\begin{figure}[h]
\includegraphics[scale=0.7]{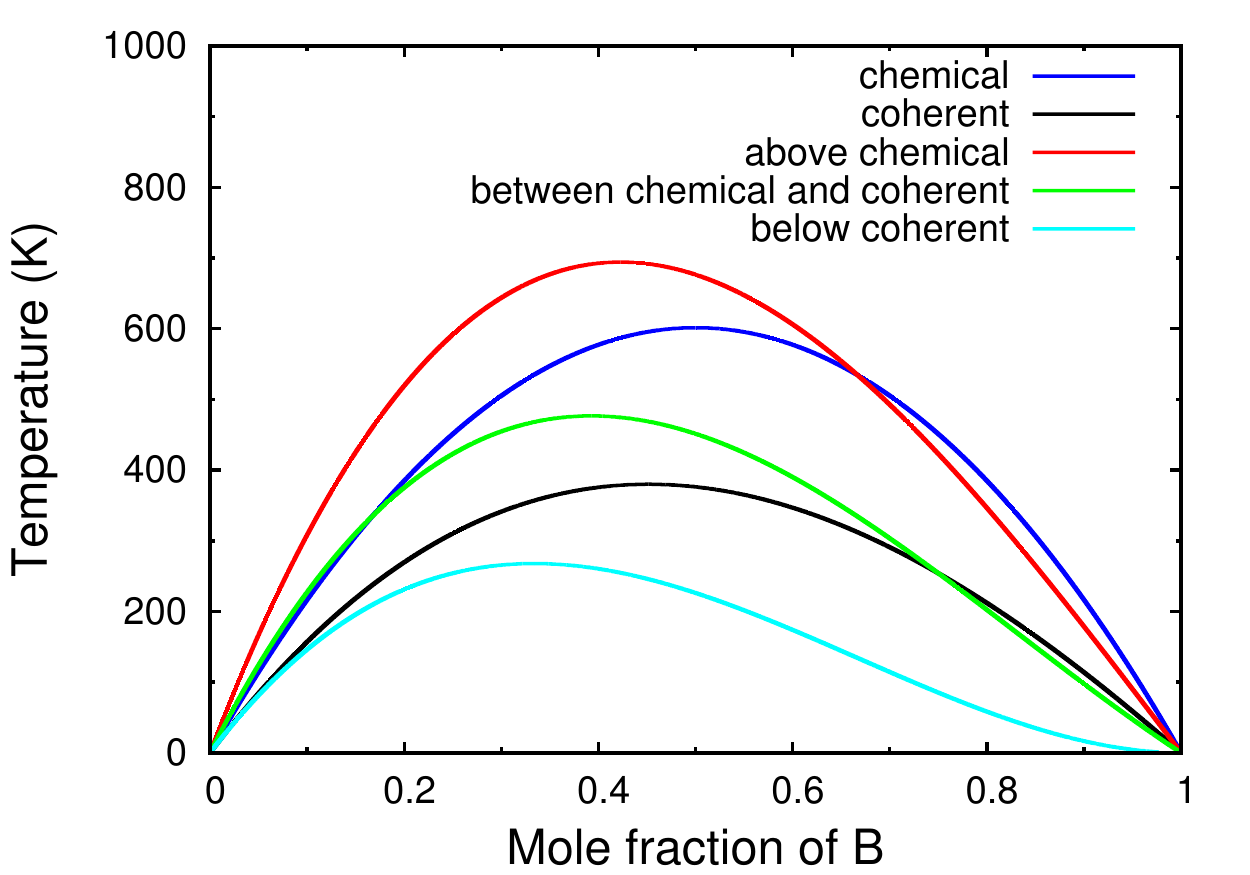}
\caption{The chemical, coherent and the constrained spinodal lines. Notice that for different values of applied homogeneous strain, the constrained spinodal can be made to lie within the coherent spinodal, or outside of coherent spinodal but within the chemical spinodal, or, even outside the chemical spinodal.}
\label{F6}
\end{figure}

The Fig.~\ref{F6}, we show the chemical, coherent and the constrained spinodal lines. The asymmetry in the spinodal lines is a characteristic of the  composition dependent modulus. The spinodal lines were constructed using the following values for the different parameters (the strains are measured with respect to a 50:50 alloy): $\Omega=10000$\,J/mol-K,  $Y_{A}=312.5$\,GPa, $Y_{B}=625$\,GPa, $\eta=0.02$, $\nu=0.3$ and $V_{A}=V_{B}=10^{-5} {\text{m}}^{3}\text{/mol}$. For obtaining the different constrained spinodals, we have applied different homogeneous strains: $e=0.04$ leads to the constrained spinodal extending beyond the chemical spinodal; 
$e=0.01$ leads to the constrained spinodal lying between the coherent and the chemical spinodal; $e=-0.02$ leads to the constrained spinodal being restricted inside the coherent spinodal. Thus, our choice clearly demonstrates the different possibilities 
for the constrained spinodal in an epitaxial systems where large tensile and compressive imposed strains
are possible.

\subsection{Particle splitting}

Wang et al~\cite{WangChenKhach1993} used phase field modelling to show particle splitting which 
was achieved by the nucleation of matrix phase at the centre of the precipitate. However, 
in this model, the elastic moduli of both the phases is assumed to be the
same and the elastic energy per unit volume was varied. This is not very realistic. 
On the other hand, Wang and Khachaturyan~\cite{WangKhachaturyan1995}, using the same homogeneous 
moduli approximation,  showed that high elastic anisotropy leads to cuboidal precipitates, which, 
due to the presence of corners that promote ``earing'' (due to the point effect of diffusion), 
can lead to star shaped precipitates; however, the stars never split (though, initial star 
shaped particles studied using sharp interface models (level set method) have shown morphologies 
closer to split morphologies: see for example, Zhao et al~\cite{ZhaoEtAl} 
-- albeit assuming inhomogeneous elasticity. 

Luo et al~\cite{LuoEtAl2007} reported particle splitting like morphologies as due to 
nucleation of ordered precipitates at dislocations (assuming homogeneous moduli approximation).
Similarly, the phase field study of Banerjee et al~\cite{BanerjeeEtAl1999} had shown that
it is possible to obtain experimentally seen splitting morphologies through particle coalescence 
(as also experimentally shown by~\cite{Calderon1,Calderon2,KisielowskiEtAl2007}).

There are other phase field models that report splitting (in elastically inhomogeneous systems). 
Li and Chen~\cite{LiChen1999} report particle splitting in systems with applied stresses.
Boussinot et al~\cite{BoussinotEtAl2010} attribute particle
splitting to elongated particles in an unfavourable direction under applied stress.
Similar conclusions are also drawn by Lee~\cite{Lee1996} using DAM method and 
Leo et al~\cite{LeoLowengrubNie2001} using a sharp 
interface model; our own work on particle 
splitting using phase field modelling also supports this conclusion~\cite{GuruThesis}.
 
Cha et al~\cite{ChaEtAl} and Kim et al~\cite{KimEtAl2007} report splitting as due to elastic 
anisotropy and diffusion induced instability (which, in some sense, is closer to ATG instability since they
also assume elastic inhomogeneity): The elastic anisotropy leads to cuboidal precipitates; the corners 
lead to earing of the precipitates as they grow; this earing enhances the elastic stress fields 
(like in ATG instability) and hence leads to splitting.

Zhu et al~\cite{ZhuChenShen2001} argue that 
splitting during coarsening is due to the geometric aspect of high aspect ratios of 
length to width of the particles. 

Leo et al~\cite{LeoLowengrubNie2001} have used sharp interface model to show that 
deviatoric applied stresses and non-dilational misfit 
strains (in the absence of applied stresses) can lead to particle splitting. 
Unfortunately, as far as we know, there are no phase field 
models that report splitting in systems with non-dilatational misfit or 
deviatoric applied strains. This would be an interesting problem
that can be solved with existing implementations of phase field models.  

Further, Lee~\cite{Lee2000} classifies elastic splitting instability into two types, namely, commensurate and 
incommensurate; incommensurate instability is the instability due to the elastic anisotropy of the matrix and 
precipitate phases being of opposite signs. Lee has simulated both types
of splitting using Discrete Atom Method and has indicated that splitting can happen even in 
elastically isotropic systems~\cite{Lee1996,Lee1997,Lee1998,ChoyLee2000}. There are no detailed studies 
on incommensurate splitting instability 
using phase field models -- though, this is again a problem that can be easily studied 
using available phase field models.

To summarise, currently, there are at least two valid mechanisms by which `split'-like patterns can
be formed. In inhomogeneous systems, it is the interactions of anisotropy induced geometries interacting
with diffusion fields leading to stress fields that result in actual splitting (through an ATG like mechanism). 
The second one is the coalescence of different ordered domains coming together during coarsening. 
Note that while the first mechanism necessarily involves elastic inhomogeneities, the second can operate
even in homogeneous systems (and while the first one is a true elastic stress induced instability, the second 
one is not). Finally, the applied stress fields, non-dilatational eigenstrains and differences in elastic anisotropy
between the matrix and precipitate phases can also 
have a strong say on splitting -- though they are not explored experimentally enough (nor by modelling in an
exhaustive manner) at the moment.

Our foregoing discussion is also very instructive at another level. 
It clearly shows that several different mechanisms can lead to the same microstructural feature. 
It also shows that phase field models (or any modelling study for that matter) can not only be used to 
verify a proposed mechanism, but also for advancing new mechanisms which can then be checked through 
experiments. Thus, while `equations without a phenomenological  background remain a formal 
game'~\cite{Nozieres2012}, these games can be very 
fruitful if they lead to such experimental validations and verifications. However, such verifications
also imply that the parameters used in simulations are realistic; checking that indeed all the parameters used 
in the simulations are realistic becomes difficult due to the different non-dimensionalisations used. Hence, 
indicating to the readers the translation of simulation parameters in terms of what they correspond to in real life 
(which, unfortunately, is not the current practice) will make the simulation studies more grounded in 
phenomenology.

\subsection{Rafting}

There have been several papers questioning elastic energy based 
explanations of rafting. For example, one of the conclusions of 
Ichitsubo et al~\cite{IchitsuboEtAl2003} reads as follows:
\begin{quote}
In the coherent elastic regime, the rafted structure cannot be realized unless the elastic misfit
exists, and both signs of lattice misfit and external stress are not relevant to the choice
of the rafted structures; the only 0 0 1 rafted structure can be formed in any conditions. This
indicates that the actual rafting phenomena cannot be explained within the elastic regime.
\end{quote}
There are also claims that plastic prestrain is essential for rafting. 
For example, Tinga et al~\cite{TingaEtAl2009} write:
\begin{quote}
Whereas a certain amount of plastic deformation is a requirement for the onset of rafting, the presence of an external
stress surprisingly appears not to be a requirement to sustaining the rafting process.
\end{quote}
Finally, there are studied based on elastic energy calculations that the raft structure itself is elastically unstable~\cite{TanakaEtAl2008} and phase field models to simulate the collapse of the rafted 
structure~\cite{TsukadaEtAl2008}. 
 
\begin{figure}[thpb]
\begin{center}
\resizebox{2in}{!}{\rotatebox{0}
{\includegraphics[clip,trim=6cm 6cm 6cm 6cm, width=1.0\textwidth]{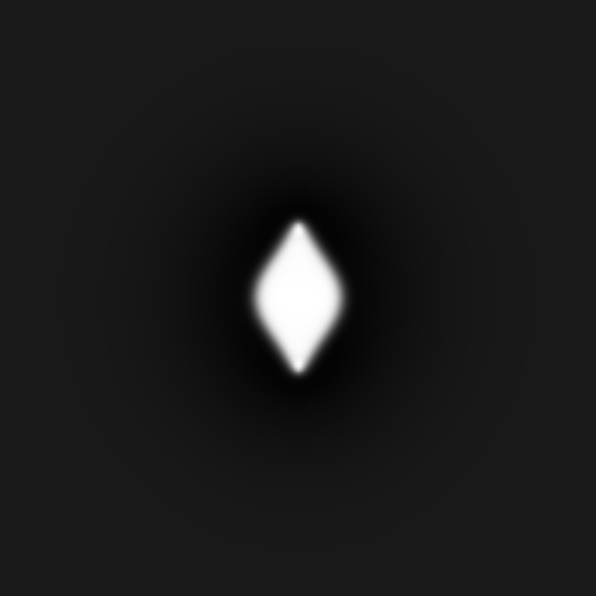}}}
\caption{Figure from~\cite{GuruThesis}: a very soft precipitate ($\delta = 0.01$) under compressive
stress ($\sigma^{A} = -0.01 G^{m}$ along $x$-axis) in an elastically isotropic system after 
300 time units; numerical simulation corresponding 
to region 5 of the Figure~\ref{SG_schematic}. } 
\label{region5}
\end{center}
\end{figure}

\begin{figure}[thpb]
\begin{center}
\subfigure[]{{
\resizebox{2in}{!}{\rotatebox{0}
{\includegraphics{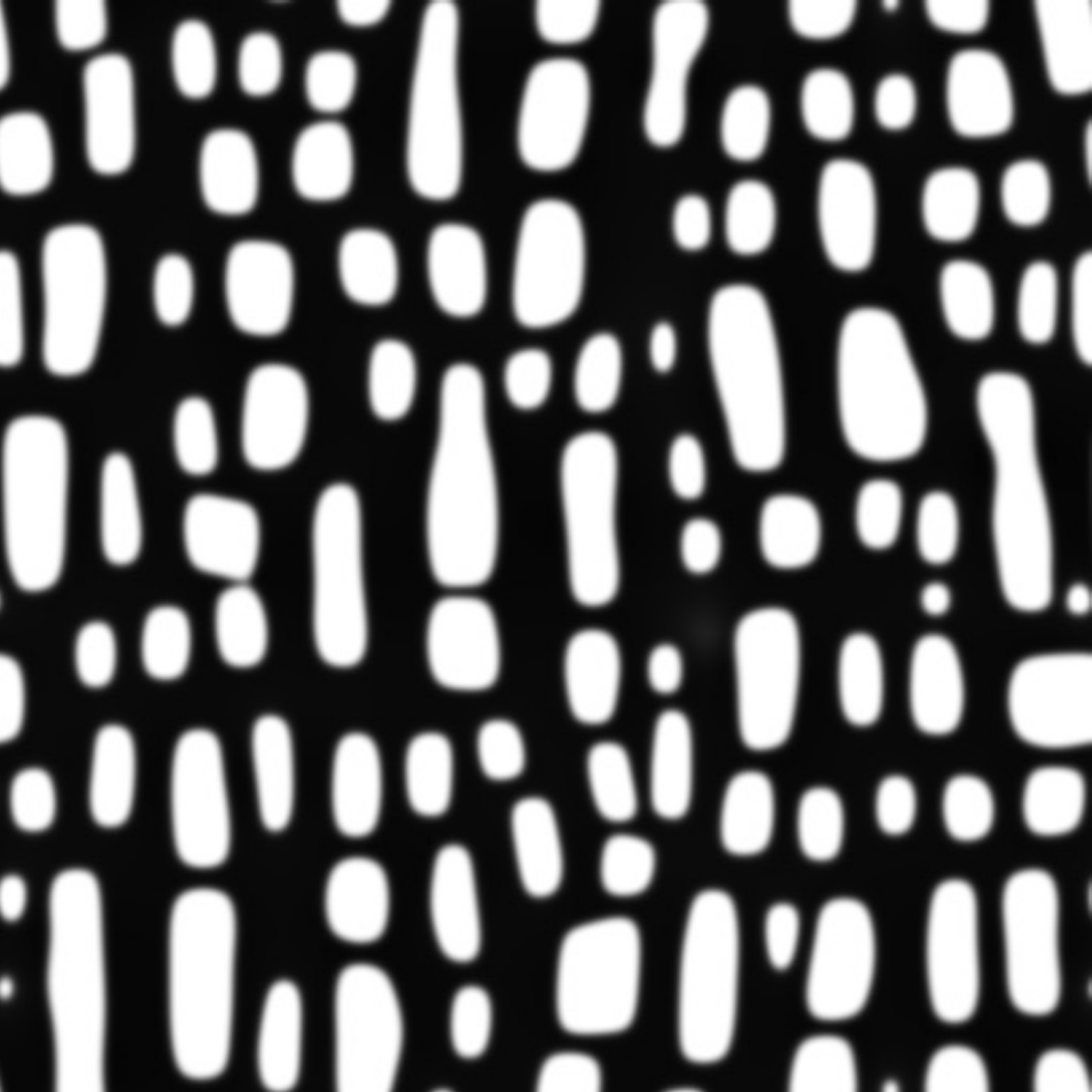}}}}}
\subfigure[]{{
\resizebox{2in}{!}{\rotatebox{0}
{\includegraphics{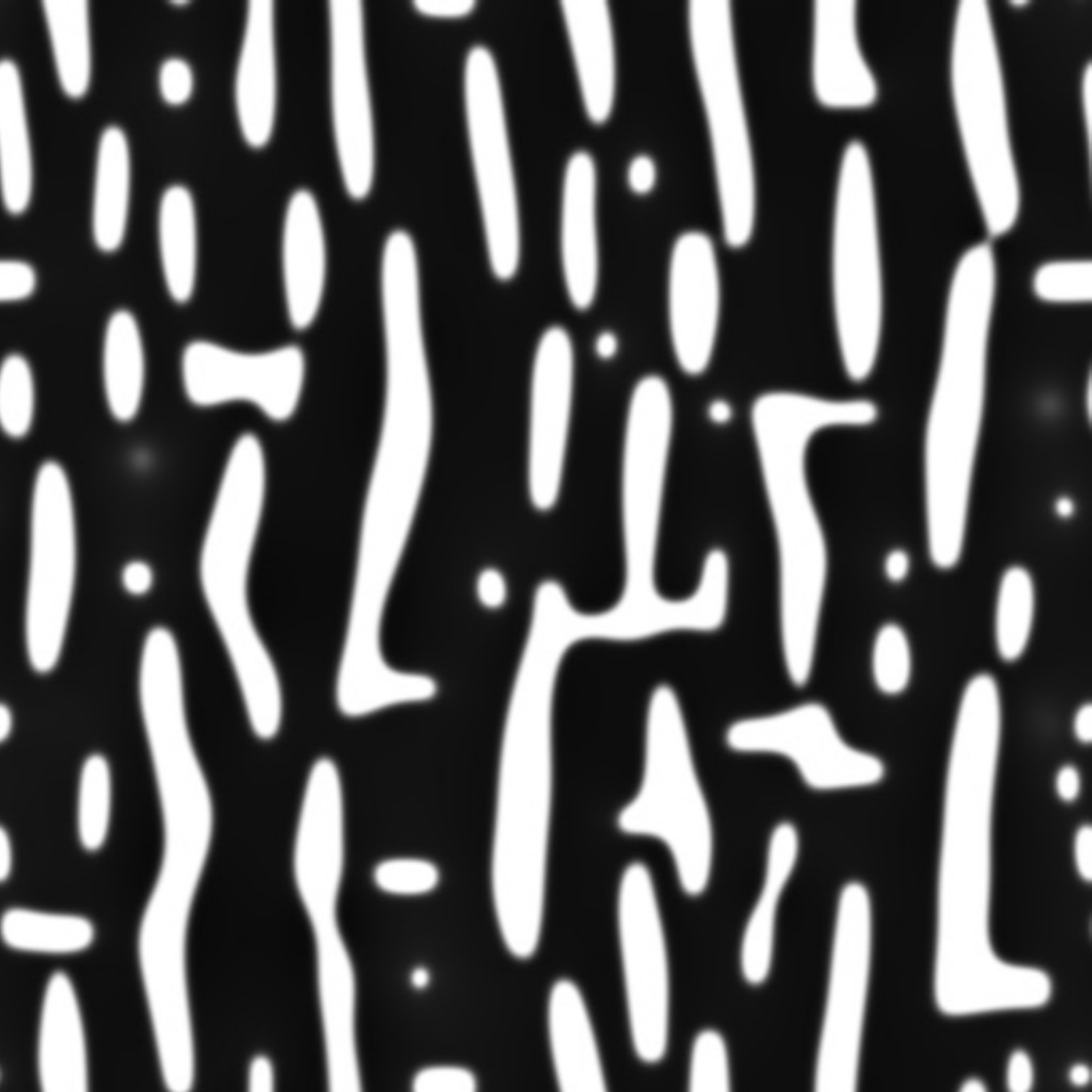}}}}}
\caption{Figures from~\cite{GuruThesis}. 
Rafting in an anisotropic system (Zener anisotropy parameter: $A_Z = 3$) of 
(a) hard particles ($\delta = 2$) under a tensile stress, and, 
(b) soft particles ($\delta = 0.5$) under a compressive stress;
stresses are applied along the $y$-axis and, the magnitude of the stress is 1\% of the shear
modulus of the matrix.
microstructures after 3000 time units. }
\label{aniso_raft}
\end{center}
\end{figure}

The most important contribution from phase field modelling is to show that purely elastic stress
driven rafting is possible. There have been a series of phase field studies showing purely elastic stress
driven rafting: Li and Chen~\cite{LiChen1997,LiChen1998}, Leo et al~\cite{LeoLowengrubNie2001}, Zhu et al~\cite{ZhuChenShen2001}, Gururajan and Abinandanan~\cite{GuruAbiRafting2007} and 
Boussinot et al~\cite{BoussinotEtAl2010}. In Fig.~\ref{region5}, and~\ref{aniso_raft}, we
show examples of purely elastic stress driven rafting~\cite{GuruThesis}; these figures
vouch for the correctness of the predictions of Schmidt and Gross~\cite{SchmidtGross}. 

Of the elastic stress driven rafting studies, Boussinot et al~\cite{BoussinotEtAl2010} is 
the most complete; it not only includes the compositional order parameter but also the 
non-conserved order parameters to account
for the different variants of of the precipitate phase. These studies show the correctness of the 
thermodynamic models based on Hookean elasticity.

As in the case of incommensurate splitting shown by Lee~\cite{Lee2000}, incommensurate
rafting is also possible~\cite{SchmidtGross}. However, a detailed study of the rafting
due to such differences in anisotropy and Poisson's ratio (which is well within the capabilities
of the current formulations and implementations) has not yet been made.

The second important contribution of the phase field models is to indicate the kinetic paths of rafting.
In this regard, the pure elasticity based phase field models described above have limited use.
However, since in the practical scenario, there is always plastic activity, and since plastic activity
can give rise to very different kinetics and kinetic paths, it becomes important to incorporate plasticity 
in the phase field models; there have been a few such improved phase field models in the last few years~\cite{GaubertEtAl2008,FinelEtAl2010}.

\subsection{ATG instabilities}

Kassner and Misbah~\cite{KasnerMisbah1999} and 
Kassner et al~\cite{KasnerEtAl2001,KasnerEtAl2001CrystalGrowth} have studied ATG instabilities by
modelling a stressed solid in contact with its melt. Kassner et al~\cite{KasnerEtAl2001} show that the
reference state to measure displacements (and hence strain and stress) is important and different choices
lead to different evolution equations; they also show that the model, in the sharp interface limit, recovers
the continuum equations for ATG instabilities. Further, Kassner et al also show that phase field models themselves
can be used to build more complex sharp interface models. 

Phase field models of ATG instabilities in the case of films in contact with vapour or vacuum is more common; see
for example~\cite{MullerGrant1999,WiseEtAl2004,WangEtAl2004,WiseEtAl2005,
Paret2005,SeolEtAl2005,NiEtAl2005,YeonEtAl2006}. In addition,
there are also phase field models that study ATG in the dynamic setting of growing films; see for 
example~\cite{EgglestonThesis,EgglestonVoorhees2002,RatzEtAl2006,KimEtAl2006,PangHuang,TakakiEtAl2008}. In addition to surface diffusion being a relatively faster
process, an important physics associated with these problems is the interfacial anisotropy and some of these
models do incorporate interfacial energy anisotropy~\cite{EgglestonVoorhees2002,WiseEtAl2004}. Since typical 
analytical studies of ATG 
instabilities assume isotropic interfacial energies~\cite{SridharRickmanSrolovitz1997}, phase field studies are helpful to
relax this assumption and see the effect of the same.

\begin{figure}
\begin{center}
\subfigure[]{{
\resizebox{2in}{!}{\rotatebox{0}
{\includegraphics{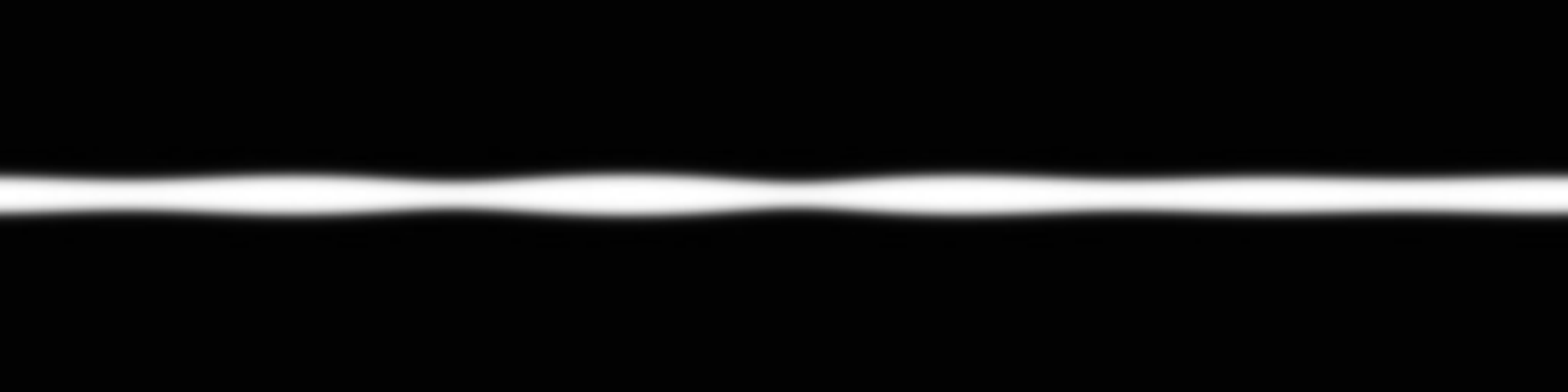}}}}}
\subfigure[]{{
\resizebox{2in}{!}{\rotatebox{0}
{\includegraphics{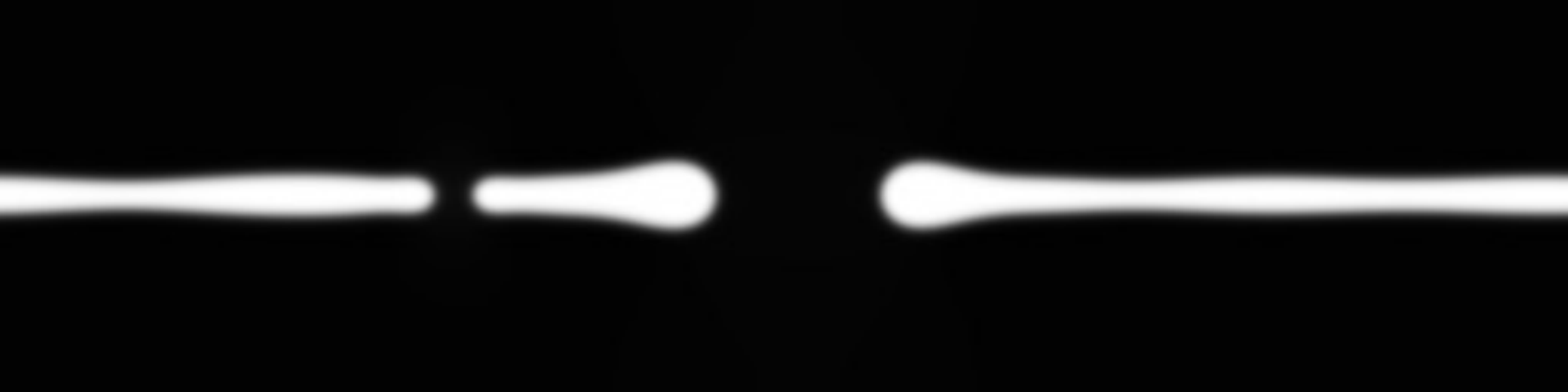}}}}}
\subfigure[]{{
\resizebox{2in}{!}{\rotatebox{0}
{\includegraphics{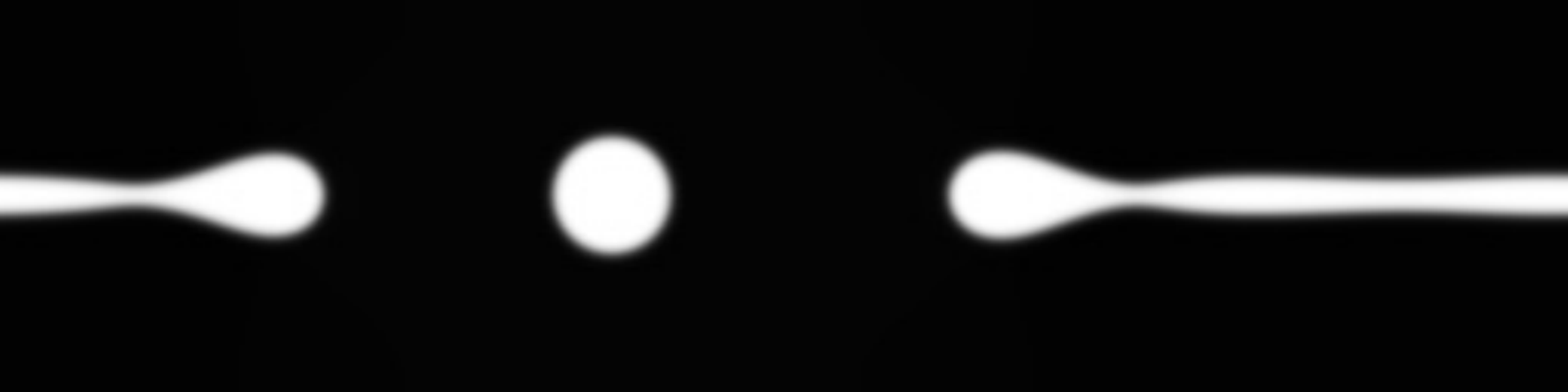}}}}}
\subfigure[]{{
\resizebox{2in}{!}{\rotatebox{0}
{\includegraphics{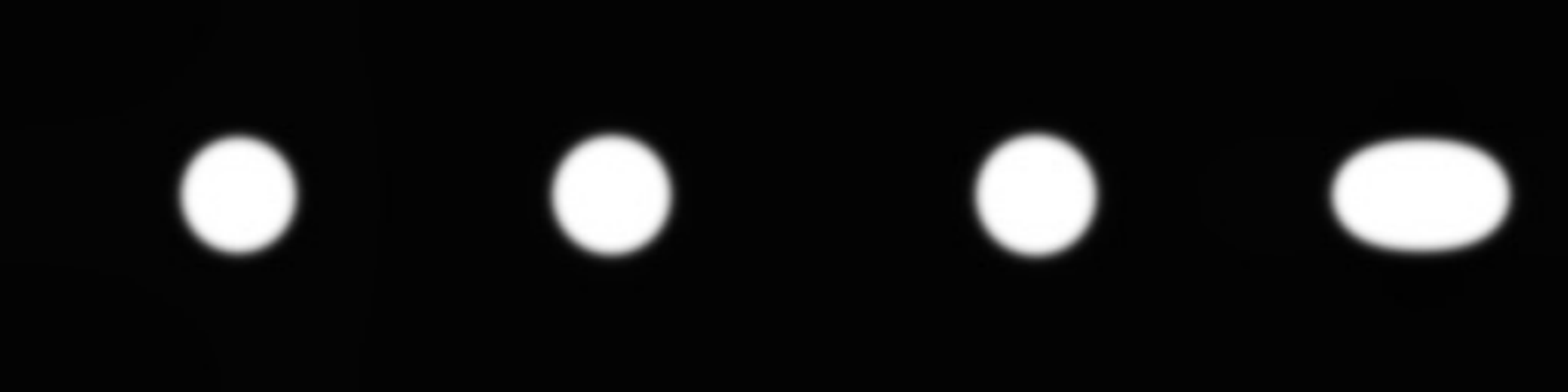}}}}}
\caption{Figures from~\cite{GuruThesis}. Symmetric break-up and late stage evolution
in a thin film assembly in an elastically isotropic, inhomogeneous ($\delta=2$) system: 
morphology at (a)~115000, (b)~122000, (c)~129000, and (d)~143000 
time units. $L_x = 512$; $L_y = 128$; $h = 10$.}
\label{sym}
\end{center}
\end{figure}

\begin{figure}
\begin{center}
\subfigure[]{{
\resizebox{3in}{!}{\rotatebox{0}
{\includegraphics{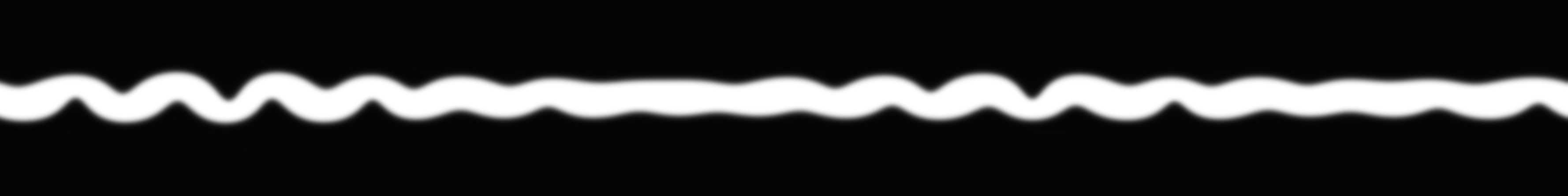}}}}}
\subfigure[]{{
\resizebox{3in}{!}{\rotatebox{0}
{\includegraphics{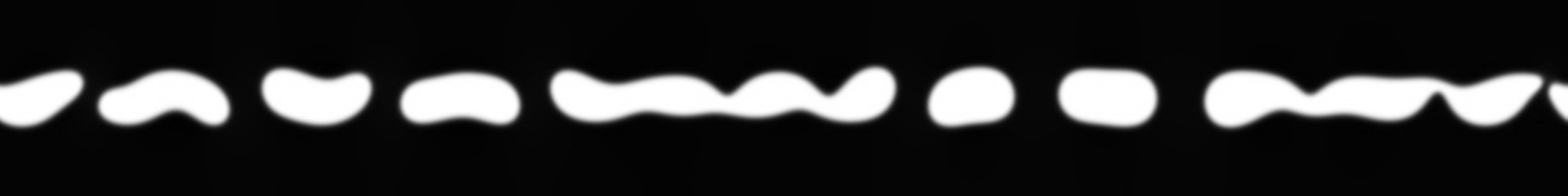}}}}}
\subfigure[]{{
\resizebox{3in}{!}{\rotatebox{0}
{\includegraphics{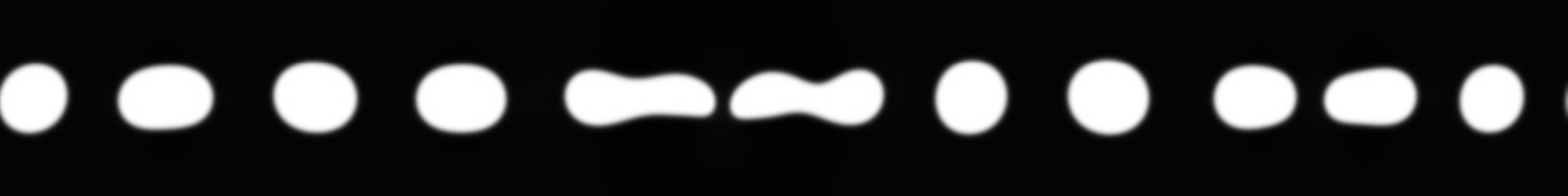}}}}}
\subfigure[]{{
\resizebox{3in}{!}{\rotatebox{0}
{\includegraphics{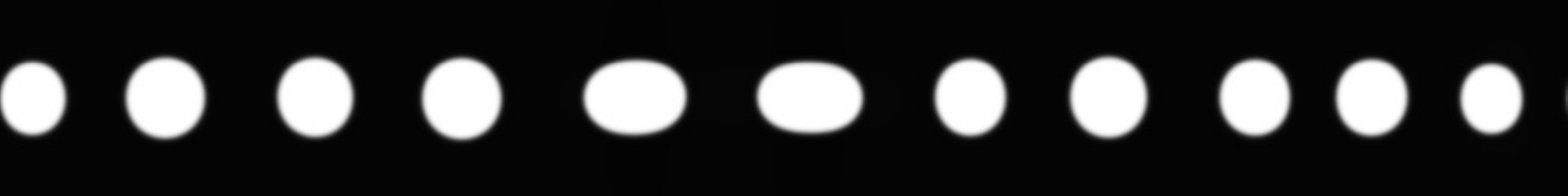}}}}}
\caption{Figures from~\cite{GuruThesis}. Anti-symmetric break-up and late stage evolution
in a thin film assembly in an elastically isotropic, inhomogeneous ($\delta=4$) system: 
morphology at (a)~20000, (b)~25000, (c)~28000, and (d)~34000 time units.
$L_x = 1024$; $L_y = 128$; $h = 20$.}
\label{asym}
\end{center}
\end{figure}

Phase field models of ATG instabilities in the cases of film assemblies are studied by 
Chirranjeevi et al~\cite{ChirranjeeviEtAl}, and Zaeem and Mesarovic~\cite{ZaemMesarovic2011}. As noted in the previous
section, in the case of solid-solid ATG instabilities, there are more than one mode of break-up, namely,
symmetric and anti-symmetric are possible. Phase field models are able to capture these different modes
of break-up under appropriate conditions. Further, using the phase field models, it is also possible 
to study long-time dynamics (which could be, and indeed is, very different from early stage dynamics --
see Figs.~\ref{sym} and~\ref{asym}). Finally, it is known that the effect of interaction between the different layers
is important~\cite{DanescuGrenet2003,HuangDesai2003} and phase field models are capable of 
dealing with them as well as the effect of elastic anisotropies~\cite{ChirranjeeviEtAl}.

Zaeem and Mesarovic~\cite{ZaemMesarovic2011} extend the study further and look at the effect of 
metastable intermediate phase. 
In addition, they show that there is homogenisation for very thin layers. However, as we have seen in the 
spinodal section, in such films, the homogenised region could be a phase separated region albeit with a
morphology different from thin films. This problem has not yet been explored in detail even though
it is well within the capabilities of current models and implementations.  

\section{Conclusions} \label{Conclusions}

In this review, using stress effects on spinodal phase separation, particle splitting, rafting and ATG instabilities,
we have shown that phase field models are quite successful in the study of elastic stress induced microstructural 
instabilities. In some cases, such as rafting, they have acted as computer experiments to access regimes which are
experimentally well near impossible to access. In some cases, such as particle splitting, phase field models
have shown that there could be more than one mechanism leading to the observed microstructural features. In the
case of ATG instabilities and spinodal phase separation, they are very helpful to understand some of
the experimental observations; this understanding can be translated into better control of phenomena.
We have also identified some problems which can be tackled with the current phase field formulations and 
numerical implementations but have not yet been studied in detail.

From the review, it is clear that there has to be more attempts to connect phase field parameters either
with phenomenological data or with data from other methods (such as first principles and atomistic simulations)
in order to make phase field studies more quantitative. We found very few papers tackling such problems.
There are also relatively lesser number of studies on 3-D systems. We believe that the GPU based computing
will change that and we may expect more 3-D studies in the next few years.

We have also come across several attempts towards
extending the models to incorporate interfacial energy anisotropy and different aspects of plasticity; 
see for example the study on stressed incoherent solid-solid interfaces by Paret~\cite{Paret}, 
effect of coupling of defects such as dislocations, 
coherent interfaces, vacancy and interstitial discs on microstructural
evolution~\cite{ShenEtAl2008}, spinodal phase separation induced by irradiation in the presence of dislocations~\cite{HoytHaataja2011}, phase separation coupled with large elastic and large elastic-plastic deformations during Lithiation in Li-ion battery electrodes~\cite{Anand2012,DiLeoEtAl2014},
phase field modelling misfit accommodation by considering a precipitate growing into a finite 
elastic-perfectly plastic matrix~\cite{SongLiu2016}, and, phase field model to study the stability of ordered array of islands by considering the elastic interaction between them~\cite{NiEtAl2004}. In the near future, we expect
that there will be more studies incorporating interfacial anisotropies and plasticity.

There are also attempts to study strain gradient theories of Eshelby problem formulation~\cite{GaoMa2010}; we believe that 
phase field formulation of such problems will be very interesting. There are a few attempts
in this direction: see for example, the strain gradient models 
(to introduce characteristic length scales of microstructure to study  mechanical behaviour) 
based phase field modelling to study strongly elastoviscoplastic systems~\cite{ForestEtAl2011}. Finally,
the studies which include other physics such as electric and magnetic fields along with elasticity
will also become more important as is clear from several recent attempts to study electrochemical processes such as corrosion
and studies such as the phase field model for morphological evolution of vesicles in electric fields~\cite{GaoEtAl2009}.

To conclude, even though the study of elastic stress effects on material behaviour started by Gibbs about 
140 years ago, their study using phase field models is still open; and, 
the open problems span the entire range -- from theoretical formulations, to numerical
implementations to materials science concepts -- not to mention experimental verifications 
and validation.

\section*{Acknowledgements}

We thank Prof. Abinandanan, Prof. Karthikeyan, Prof. A Choudhury and Prof. D Banerjee 
of Indian Institute of Science, 
Dr R Sankarasubramanian of Defence Metallurgical Research Laboratory,  
Prof. E Chandrasekhar and Prof. Mira Mitra, Indian Institute of Technology Bombay,
Prof. S Chatterjee and S Bhattacharya of Indian Institute of Technology Hyderabad,
Prof. R Mukherjee, Indian Institute of Technology Kanpur,
Dr. Chirranjeevi Gopal, Stanford University,
Prof. Ferdinand Haider, University of Augsburg,
Prof. P Voorhees, Northwestern University,
Prof. Kuo-An Wu, National Tsing Hua University,
and Prof. G Phanikumar of Indian Institute of Technology Madras, for useful discussions.
One of us (MPG) would like to thank IRCC, IIT Bombay, DST, Government of India and Tata Steel
for financial support.

\bibliography{References}
\end{document}